\documentclass[12pt]{article}
\usepackage{epsfig}
\begin{document}
\title{Regarding the possibility to observe the LHCb hidden-charm strange pentaquark
$P_{cs}(4459)^0$ in antikaon-induced $J/\psi$ meson production on protons and nuclei near the
$J/\psi$$\Lambda$ production threshold}
\author{E. Ya. Paryev \\
{\it Institute for Nuclear Research, Russian Academy of Sciences,}\\
{\it Moscow 117312, Russia}}

\renewcommand{\today}{}
\maketitle

\begin{abstract}
We study the near-threshold $J/\psi$ meson production on protons and nuclei
by considering incoherent direct non-resonant $K^-p \to {J/\psi}\Lambda$ and two-step resonant
${K^-}p \to P_{cs}(4459)^0 \to {J/\psi}\Lambda$ charmonium production processes.
We calculate the absolute excitation functions, energy and momentum distributions
for the non-resonant, resonant and for the combined (non-resonant plus resonant) production
of $J/\psi$ mesons off protons as well as off carbon and tungsten target nuclei at near-threshold
incident antikaon energies by considering these elementary production channels as well as by
assuming the spin-parity assignment of the hidden-charm resonance $P_{cs}(4459)^0$ with strangeness
as $J^P=(3/2)^-$ within six different scenarios for the branching ratio $Br[P_{cs}(4459)^0 \to {J/\psi}\Lambda]$
of the decay $P_{cs}(4459)^0 \to {J/\psi}\Lambda$.
We show that the combined observables considered reveal definite sensitivity to these scenarios, which
means that they may be an important tool to provide further evidence for the existence of the pentaquark
$P_{cs}(4459)^0$ resonance and to get valuable information on its decay rate to the ${J/\psi}\Lambda$ final state.
Their measurements could be performed in the future at the J-PARC Hadron Experimental Facility.
\end{abstract}

\newpage

\section*{1. Introduction}
The study of the exotic hadronic states, the hidden-charm pentaquarks has received considerable interest in recent
years (see, for example, Refs. [1--7]) and becomes a hot topic after the discovery by the LHCb Collaboration
pentaquark resonances $P_c(4380)^+$ and $P_c(4450)^+$ in the ${J/\psi}p$  invariant
mass spectrum of the $\Lambda^0_b \to K^-({J/\psi}p)$ decays [8] and, especially, after the observation
by the Collaboration of three new narrow structures $P_c(4312)^+$, $P_c(4440)^+$ and $P_c(4457)^+$ in these
decays [9], based on additional collected data and on an improved selection strategy, instead of initially
claimed $P_c(4380)^+$ and $P_c(4450)^+$ states.
The quark structure of the above pentaquarks is $|P^+_c>=|uudc{\bar c}>$, i.e., they are composed of three
light quarks $u$, $u$, $d$ and a charm-anticharm pair $c{\bar c}$.
In a molecular scenario, due to the closeness of the observed
$P_c(4312)^+$ and $P_c(4440)^+$, $P_c(4457)^+$ masses to the ${\Sigma^+_c}{\bar D}^0$ and
${\Sigma^+_c}{\bar D}^{*0}$ thresholds, the $P_c(4312)^+$
resonance can be, in particular, considered as an s-wave ${\Sigma^+_c}{\bar D}^0$ bound state, while the
$P_c(4440)^+$ and $P_c(4457)^+$ as s-wave ${\Sigma^+_c}{\bar D}^{*0}$ bound molecular states [10--23].
More recently, the LHCb Collaboration discovered a new narrow hidden-charm pentaquark with strangeness denoted as
$P_{cs}(4459)^0$ in the invariant mass spectrum of the ${J/\psi}\Lambda$ in the
$\Xi^-_b \to K^-({J/\psi}\Lambda)$ decays [24]. Only the mass and total width of the $P_{cs}(4459)^0$
were measured [24], while its spin-parity quantum numbers, decay rates and the internal structure
(loosely-bound hadronic molecular state or tightly-bound compact pentaquark state) are still unknown.
Since the $P_{cs}(4459)^0$ state is just below the $\Xi_c{\bar D}^*$ threshold,
it is natural to interpret it as the hidden-charm strange $\Xi_c{\bar D}^*$ molecule (see, for instance,
Ref. [25] and those given below). However, there exists other explanation [26] of the $P_{cs}$ states,
based on the quark model accounting for their $|udsc{\bar c}>$ valence quark content.
The existence of hidden-charm strange pentaquark resonances $P_{cs}$,
which are a strange counterparts of the $P_c$ states, has been predicted
a few years before the LHCb observation [24] in some earlier papers (see, for example, [1, 2, 27--30]).
Furthermore, these resonances were suggested to be searched for in the
$\Xi^-_b \to {K^-}({J/\psi}\Lambda)$ [31, 32], $\Lambda_b \to {\eta}({J/\psi}\Lambda)$ [33],
$\Lambda_b \to {K^0}({J/\psi}\Lambda)$ [34] and $\Lambda_b \to {\phi}({J/\psi}\Lambda)$ [35] decays.
It is worth noting that another hidden-charm exotic state with strangeness -- the strange hidden-charm
tetraquark state $Z_{cs}(3985)^-$ has been observed very recently by the BESIII Collaboration in the processes
$e^+e^- \to K^+(D_s^-D^{*0}+D_s^{*-}D^0)$ [36]. This state has the minimum quark content
$Z_{cs}(3985)^-=|{\bar u}sc{\bar c}>$. It is in the proximity of the $D_s^-D^{*0}/D_s^{*-}D^0$
mass thresholds and can be interpreted [37, 38] as strange molecular partner
with hadronic molecular configuration $D_s^-D^{*0}-D_s^{*-}D^0$
of the non-strange tetraquark resonance $Z_{c}(3900)^-$, having the valence quark content
$Z_{c}(3900)^-=|{\bar u}dc{\bar c}>$
and possible hadronic molecular structure $D^-D^{*0}-D^{*-}D^0$.
In addition, one needs to note that possible hidden-charm molecular pentaquark states
with double and triple strangeness have been investigated in Refs. [39] and [40], respectively.

 To understand better the $P_{cs}(4459)^0$ state, observed in the $\Xi^-_b$ decay at LHCb, it is also
of importance to investigate its production in other possible reactions. For example, the energy of the
negative kaon beam, which will be available at the K10 beam line in the extended J-PARC Hadron Experimental Facility
[41, 42], will be sufficient to observe the $P_{cs}(4459)^0$ pentaquark during the $K^-p \to {J/\psi}\Lambda$ process.
In fact, in Ref. [43] the contribution of this pentaquark to the elementary $K^-p \to {J/\psi}\Lambda$ reaction
has been determined, employing two different theoretical approaches, i.e., the effective Lagrangian and the
Regge models. It was shown that the $P_{cs}(4459)^0$ can be also searched for through
a scan of the total and differential cross sections of this reaction.

 In the present study, we consider the contribution of the $P_{cs}(4459)^0$ state to
charmonium $J/\psi$ production by $K^-$ mesons on protons and nuclei near threshold by using
the standard Breit-Wigner prescription for this contribution and by employing the available
scarce experimental information on the total cross sections of the $K^-p \to {J/\psi}X$ and
$K^-p \to {\phi}X$ processes to estimate the background contribution.
The consideration is based on the model developed in Ref. [44] and devoted to the study the role
of the pentaquark resonance $P_c(4450)^+$ in $J/\psi$ photoproduction on nuclei at near-threshold
incident photon energies of 5--11 GeV.
We briefly recapitulate the main assumptions of the model [44] and describe, where
necessary, the corresponding extensions. Additionally, we present the predictions obtained within this
expanded model for the $J/\psi$ excitation functions, energy and momentum distributions
in ${K^-}p$ as well as in ${K^-}$$^{12}$C and ${K^-}$$^{184}$W collisions at
near-threshold incident energies. These predictions may serve as guidance for future dedicated experiment
at the J-PARC facility.

\section*{2. The model}

\subsection*{2.1. Direct non-resonant $J/\psi$ production mechanism}

  Direct non-resonant production of $J/\psi$ mesons in antikaon-nucleus reactions in the near-threshold center-of-mass
$K^-$ beam energy region 4.2126 GeV $\le$ $\sqrt{s}$ $\le$ 4.6126 GeV
\footnote{$^)$In which the mass of the observed [24] hidden-charm strange pentaquark state $P_{cs}(4459)^0$ is concentrated and where it can be observed [43] in the $K^-p$ reactions.}$^)$
,
corresponding to the excess energies $\epsilon_{{J/\psi}\Lambda}$ above the lowest ${J/\psi}\Lambda$ production
threshold $\sqrt{s_{\rm th}}=m_{J/\psi}+m_{\Lambda}=4.2126$ GeV ($m_{J/\psi}$ and $m_{\Lambda}$ are the $J/\psi$ meson and $\Lambda$ hyperon bare masses, respectively),
0 $\le$ $\epsilon_{{J/\psi}\Lambda}=\sqrt{s}-\sqrt{s_{\rm th}}$ $\le$ 0.4 GeV,
or to the laboratory incident $K^-$ beam momenta 8.844 GeV/c $\le$ $p_{K^-}$ $\le$ 10.728 GeV/c
\footnote{$^)$Which are well in the near future within the capabilities  of the planned K10 beam line at
the J-PARC Hadron Experimental Facility [41, 42].}$^)$
,
may occur in the following $K^-p$ elementary process, which has, respectively, the lowest free $J/\psi$
production threshold momentum (8.844 GeV/c):
\begin{equation}
{K^-}+p \to J/\psi+\Lambda.
\end{equation}
We can neglect in the incident momentum range of interest the contribution to the $J/\psi$ yield
from the processes $K^-N \to {J/\psi}\Sigma$, $K^-p \to {J/\psi}{\Lambda}\pi^0$
due to larger their production threshold momenta ($\approx$ 9.19 and 9.46 GeV/c, respectively)
in free ${K^-}N$ interactions.
In line with [45], we ignore the modification of the incoming high-momentum $K^-$ meson
mass in the nuclear matter. Furthermore, we also neglect the medium modification of the outgoing high-momentum
$J/\psi$ and $\Lambda$ (see below) masses in the present work.

Accounting for the attenuation of the incident antikaon and the final full-sized [46] $J/\psi$ meson in the nuclear matter in terms, respectively, of the $K^-N$ total cross section $\sigma_{K^-N}^{\rm tot}$ and the ${J/\psi}N$
absorption cross section $\sigma_{{J/\psi}N}$, we represent, according to Refs. [47--49],
the inclusive differential and total
\footnote{$^)$In the full allowed phase space without any cuts on the angle and momentum of the
$J/\psi$ meson.}$^)$
cross sections for the production of $J/\psi$ mesons with the
momentum ${\bf p}_{J/\psi}$ off nuclei in the direct non-resonant antikaon-induced channel (1) as follows:
\begin{equation}
\frac{d\sigma_{{K^-}A\to {J/\psi}X}^{({\rm dir})}({\bf p}_{K^-},{\bf p}_{J/\psi})}
{d{\bf p}_{J/\psi}}=\left(\frac{Z}{A}\right)
I_{V}[A,\sigma_{{J/\psi}N}]
\left<\frac{d\sigma_{{K^-}p \to {J/\psi}\Lambda}({\bf p}_{K^-},{\bf p}_{J/\psi})}{d{\bf p}_{J/\psi}}\right>_A,
\end{equation}
\begin{equation}
\sigma_{{K^-}A\to {J/\psi}X}^{({\rm dir})}({\bf p}_{K^-})=\left(\frac{Z}{A}\right)
I_{V}[A,\sigma_{{J/\psi}N}]
\left<\sigma_{{K^-}p \to {J/\psi}\Lambda}({\bf p}_{K^-})\right>_A;
\end{equation}
where
\begin{equation}
I_{V}[A,\sigma]=2{\pi}A\int\limits_{0}^{R}r_{\bot}dr_{\bot}
\int\limits_{-\sqrt{R^2-r_{\bot}^2}}^{\sqrt{R^2-r_{\bot}^2}}dz
\rho(\sqrt{r_{\bot}^2+z^2})
\end{equation}
$$
\times
\exp{\left[-A\sigma_{K^-N}^{\rm tot}\int\limits_{-\sqrt{R^2-r_{\bot}^2}}^{z}
\rho(\sqrt{r_{\bot}^2+x^2})dx
-A{\sigma}\int\limits_{z}^{\sqrt{R^2-r_{\bot}^2}}
\rho(\sqrt{r_{\bot}^2+x^2})dx\right]},
$$
\begin{equation}
\left<\frac{d\sigma_{{K^-}p \to {J/\psi}\Lambda}({\bf p}_{K^-},{\bf p}_{J/\psi})}{d{\bf p}_{J/\psi}}\right>_A=
\int\int
P_A({\bf p}_t,E)d{\bf p}_tdE
\left[\frac{d\sigma_{{K^-}p \to {J/\psi}\Lambda}(\sqrt{s^*},{\bf p}_{J/\psi})}{d{\bf p}_{J/\psi}}\right],
\end{equation}
\begin{equation}
\left<\sigma_{{K^-}p \to {J/\psi}\Lambda}({\bf p}_{K^-})\right>_A=
\int\int
P_A({\bf p}_t,E)d{\bf p}_tdE
\sigma_{{K^-}p \to {J/\psi}\Lambda}(\sqrt{s^*})
\end{equation}
and
\begin{equation}
  s^*=(E_{K^-}+E_t)^2-({\bf p}_{K^-}+{\bf p}_t)^2,
\end{equation}
\begin{equation}
   E_t=M_A-\sqrt{(-{\bf p}_t)^2+(M_{A}-m_{p}+E)^{2}}.
\end{equation}
Here, $d\sigma_{{K^-}p\to {J/\psi}\Lambda}(\sqrt{s^*},{\bf p}_{J/\psi})/d{\bf p}_{J/\psi}$
and $\sigma_{{K^-}p\to {J/\psi}\Lambda}(\sqrt{s^*})$ are, respectively, the off-shell "in-medium"
differential and total cross sections for the production of $J/\psi$ in reaction (1)
at the "in-medium" ${K^-}p$ center-of-mass energy $\sqrt{s^*}$;
$\rho({\bf r})$ and $P_A({\bf p}_t,E)$ are the local nucleon density and the nuclear
spectral function of target nucleus $A$ normalized to unity
\footnote{$^)$The specific information about these quantities, used in our calculations,
is given in Refs. [50--53].}$^)$;
${\bf p}_{K^-}$ and $E_{K^-}$ are the laboratory momentum and total
energy of the initial $K^-$ meson ($E_{K^-}=\sqrt{m_{K^-}^2+{\bf p}_{K^-}^2}$, $m_{K^-}$ is the
rest mass of a $K^-$);
${\bf p}_{t}$  and $E$ are the internal momentum and removal energy of the struck target proton
involved in the collision process (1); $Z$ and $A$ are the numbers of protons and nucleons in
the target nucleus, $M_{A}$  and $R$ are its mass and radius; $m_p$ is the bare proton mass.
For the $J/\psi$-nucleon absorption cross section $\sigma_{{J/\psi}N}$
we will employ the value $\sigma_{{J/\psi}N}=3.5$ mb
motivated by the results from the $J/\psi$ photoproduction experiment at SLAC [54, 55] (cf. [47]).
We also use $\sigma_{{K^-}N}^{\rm tot}=22$ mb in our present calculations for beam momenta of interest [56].
The quantity $(Z/A)I_{V}[A,\sigma_{{J/\psi}N}]$ in Eqs. (2), (3) represents the effective number of target
protons participating in the direct non-resonant $K^-p \to {J/\psi}\Lambda$ process. It is calculated according
to Eq. (4), in which the first and the second terms in exponent describe, respectively, the distortion of the
incident antikaon and the full-sized $J/\psi$ meson final-state absorption. It accounts for as well the fact
that in the momentum range of interest $J/\psi$ meson is produced and propagates inside the target nucleus
at small laboratory angles
\footnote{$^)$Thus, for example, at a beam momenta of 9 and 10 GeV/c the $J/\psi$ laboratory production
polar angles in reaction (1) proceeding on the free target proton being at rest are less than
2.089$^{\circ}$ and 5.452$^{\circ}$, respectively, (see below).}$^)$
without [47] quasielastic rescatterings on the intranuclear nucleons.

 Following [47--49], we assume that the off-shell "in-medium" cross sections\\
$d\sigma_{{K^-}p\to {J/\psi}\Lambda}(\sqrt{s^*},{\bf p}_{J/\psi})/d{\bf p}_{J/\psi}$
and $\sigma_{{K^-}p\to {J/\psi}\Lambda}(\sqrt{s^*})$ for $J/\psi$ production in reaction (1)
are equivalent to the respective on-shell cross sections
$d\sigma_{{K^-}p\to {J/\psi}\Lambda}(\sqrt{s},{\bf p}_{J/\psi})/d{\bf p}_{J/\psi}$ and
$\sigma_{{K^-}p \to {J/\psi}\Lambda}(\sqrt{s})$ calculated for the off-shell kinematics
of this reaction and in which, in particular, the free space center-of-mass energy squared $s$,
presented by the formula
\begin{equation}
  s=W^2=(E_{K^-}+m_p)^2-{\bf p}_{K^-}^2,
\end{equation}
is replaced by the in-medium expression (7). In our calculations, the former differential cross
section was described according to the two-body kinematics of the process (1) (cf. [48, 49]):
\begin{equation}
\frac{d\sigma_{K^{-}p \to {J/\psi}{\Lambda}}(\sqrt{s^*},{\bf p}_{J/\psi})}
{d{\bf p}_{J/\psi}}=
\frac{\pi}{I_2(s^*,m_{J/\psi},m_{\Lambda})E_{J/\psi}}
\end{equation}
$$
\times
\frac{d\sigma_{{K^{-}}p \to {J/\psi}{\Lambda}}(\sqrt{s^*},\theta_{J/\psi}^*)}{d{\bf \Omega}_{J/\psi}^*}
\frac{1}{(\omega+E_t)}\delta\left[\omega+E_t-\sqrt{m_{\Lambda}^2+({\bf Q}+{\bf p}_t)^2}\right],
$$
where
\begin{equation}
I_2(s^*,m_{J/\psi},m_{\Lambda})=\frac{\pi}{2}
\frac{\lambda(s^*,m_{J/\psi}^{2},m_{\Lambda}^{2})}{s^*},
\end{equation}
\begin{equation}
\lambda(x,y,z)=\sqrt{{\left[x-({\sqrt{y}}+{\sqrt{z}})^2\right]}{\left[x-
({\sqrt{y}}-{\sqrt{z}})^2\right]}},
\end{equation}
\begin{equation}
\omega=E_{K^-}-E_{J/\psi}, \,\,\,\,{\bf Q}={\bf p}_{K^-}-{\bf p}_{J/\psi},\,\,\,\,
E_{J/\psi}=\sqrt{m^2_{J/\psi}+{\bf p}_{J/\psi}^2}.
\end{equation}
Here, $d\sigma_{{K^{-}}p \to {J/\psi}{\Lambda}}(\sqrt{s^*},\theta_{J/\psi}^*)/d{\bf \Omega}_{J/\psi}^*$
is the off-shell differential cross section for the production of $J/\psi$ meson in reaction (1) under the
polar angle $\theta_{J/\psi}^*$ in the $K^-p$ c.m.s. This cross section is assumed to have the form
analogous to that proposed in Ref. [57] for $\phi$ mesons produced in the free space $K^-p \to {\phi}\Lambda$
reaction near the ${\phi}\Lambda$ production threshold, viz.:
\begin{equation}
\frac{d\sigma_{{K^-}p \to {J/\psi}\Lambda}(\sqrt{s^*},\theta^*_{J/\psi})}{d{\bf \Omega}_{J/\psi}^*}=
a{\rm e}^{b_{J/\psi}(t-t^+)}\sigma_{{K^-}p \to {J/\psi}\Lambda}(\sqrt{s^*}),
\end{equation}
where $t$ is the square of the 4-momentum transfer between the incident antikaon and final $J/\psi$
meson, and $t^+$ is its maximum value, corresponding to the $t$ where the $J/\psi$ is produced at
angle of 0$^{\circ}$ in the $K^-p$ c.m. frame. It can readily be expressed in terms of the total energies
and momenta of the incident $K^-$ meson ($E_{K^-}^*$ and $p_{K^-}^*$) and final $J/\psi$ meson
($E_{J/\psi}^*$ and $p_{J/\psi}^*$) in this reference frame. Indeed, we have
\begin{equation}
t=m_{K^-}^2+m_{J/\psi}^2-2E^*_{K^-}E^*_{J/\psi}+2p^*_{K^-}p^*_{J/\psi}\cos{\theta^*_{J/\psi}},
\end{equation}
where
\begin{equation}
E_{K^-}^*=\sqrt{m^2_{K^-}+p^{*2}_{K^-}}, \,\,\,\,
E_{J/\psi}^*=\sqrt{m^2_{J/\psi}+p^{*2}_{J/\psi}}
\end{equation}
and
\begin{equation}
p_{K^-}^*=\frac{1}{2\sqrt{s^*}}\lambda(s^*,m_{K^-}^2,E_{t}^2-p_t^2),
\end{equation}
\begin{equation}
p_{J/\psi}^*=\frac{1}{2\sqrt{s^*}}\lambda(s^*,m_{J/\psi}^{2},m_{\Lambda}^2).
\end{equation}
From Eq. (15), we straightforwardly get
\begin{equation}
t^+=m_{K^-}^2+m_{J/\psi}^2-2E^*_{K^-}E^*_{J/\psi}+2p^*_{K^-}p^*_{J/\psi}.
\end{equation}
Taking into consideration Eqs. (15) and (19), we can represent the quantity $t-t^+$,
entering into equation (14), in the form
\begin{equation}
t-t^+=2p^*_{K^-}p^*_{J/\psi}(\cos{\theta^*_{J/\psi}}-1).
\end{equation}
Equating expression (15) to its value in the laboratory system, one can then express the angle of
$J/\psi$ meson production in the $K^-p$ c.m. frame, $\theta^*_{J/\psi}$, via the production angle,
$\theta_{J/\psi}$, in the laboratory frame
($\cos{\theta_{J/\psi}}={\bf p}_{K^-}{\bf p}_{J/\psi}/p_{K^-}p_{J/\psi}$).
As a result, we arrive at
\begin{equation}
\cos{\theta_{J/\psi}^*}=\frac{p_{K^-}p_{J/\psi}\cos{\theta_{J/\psi}}+
(E_{K^-}^*E_{J/\psi}^*-E_{K^-}E_{J/\psi})}{p_{K^-}^*p_{J/\psi}^*}.
\end{equation}
Since the $J/\psi$ angular distribution from the reaction $K^-p \to {J/\psi}\Lambda$
is experimentally unknown and the quark structure of the $J/\psi$ vector meson
is similar to that of the $\phi$ vector meson (they, respectively,
are: $|J/\psi>=|c{\bar c}>$, $|\phi>=|s{\bar s}>$) as well as because the $J/\psi$ production on the
proton by the $K^-$ meson beam near the ${J/\psi}\Lambda$ threshold is similar to the
$\phi$ meson production near the ${\phi}\Lambda$ threshold (see Fig. 1),
we will assume that the slope parameter $b_{J/\psi}$ in Eq. (14) is the same as an exponential
$t$-slope $b_{\phi}$ of the differential cross section of the reaction $K^-p \to {\phi}\Lambda$
in the c.m. system near the threshold
\footnote{$^)$The threshold antikaon momentum for ${\phi}\Lambda$ production on a free proton being at rest is
1.762 GeV/c.}$^)$
.
The lowest incident $K^-$ beam momentum for which this cross section
has been measured [58] is ${\tilde p}_{K^-}=2.24$ GeV/c. In Ref. [57] it has been fitted
as (cf. Eqs. (14) and (20))
\begin{equation}
\frac{d\sigma_{{K^-}p \to {\phi}\Lambda}}{d{\bf \Omega}_{\phi}^*}=
{\tilde a}{\rm e}^{{\tilde {\alpha}}(\cos{\theta^*_{\phi}}-1)}.
\end{equation}
Here, $\theta_{\phi}^*$ is the $\phi$ c.m.s. production angle in the ${K^-}p \to {\phi}\Lambda$
reaction and
\begin{equation}
{\tilde {\alpha}}=2{\tilde p}^*_{K^-}p^*_{\phi}b_{\phi}=1.8,
\end{equation}
where initial antikaon momentum in the c.m. frame ${\tilde p}^*_{K^-}$ is defined now as
\begin{equation}
{\tilde p}_{K^-}^*=\frac{1}{2\sqrt{{\tilde s}}}\lambda({\tilde s},m_{K^-}^2,m_p^2)
\end{equation}
and final $\phi$ meson momentum in this frame $p^*_{\phi}$ is given by
\begin{equation}
p_{\phi}^*=\frac{1}{2\sqrt{{\tilde s}}}\lambda({\tilde s},m_{\phi}^{2},m_{\Lambda}^2).
\end{equation}
Here, $m_{\phi}$ is the bare $\phi$ meson mass.
The relation between the free collision energy squared ${\tilde s}$, entering into the expressions (24), (25),
and the total energy ${\tilde E}_{K^-}$ and momentum ${\tilde p}_{K^-}$ of the $K^-$ meson
inducing the reaction ${K^-}p \to {\phi}\Lambda$ is similar to that given above by Eq. (9).
For ${\tilde p}_{K^-}=2.24$ GeV/c we have $\sqrt{{\tilde s}}=2.3299$ GeV and
${\tilde p}^*_{K^-}=0.902$ GeV/c, $p_{\phi}^*=0.466$ GeV/c.
So, using Eq. (23), we readily obtain that $b_{\phi}=2.1$ GeV$^{-2}$. And, thus, in line with the aforementioned,
the value of the $J/\psi$ slope parameter $b_{J/\psi}$ in Eq. (14) is $b_{J/\psi}=2.1$ GeV$^{-2}$.
It is interesting to note that this value is close to the $t$-slope of the differential cross section
of the reaction ${\gamma}p \to {J/\psi}p$ of $\approx$1.67 GeV$^{-2}$, measured recently at GlueX [59]
in the near-threshold energy region. We will employ it in our subsequent cross-section calculations.
\begin{figure}[htb]
\begin{center}
\includegraphics[width=14.0cm]{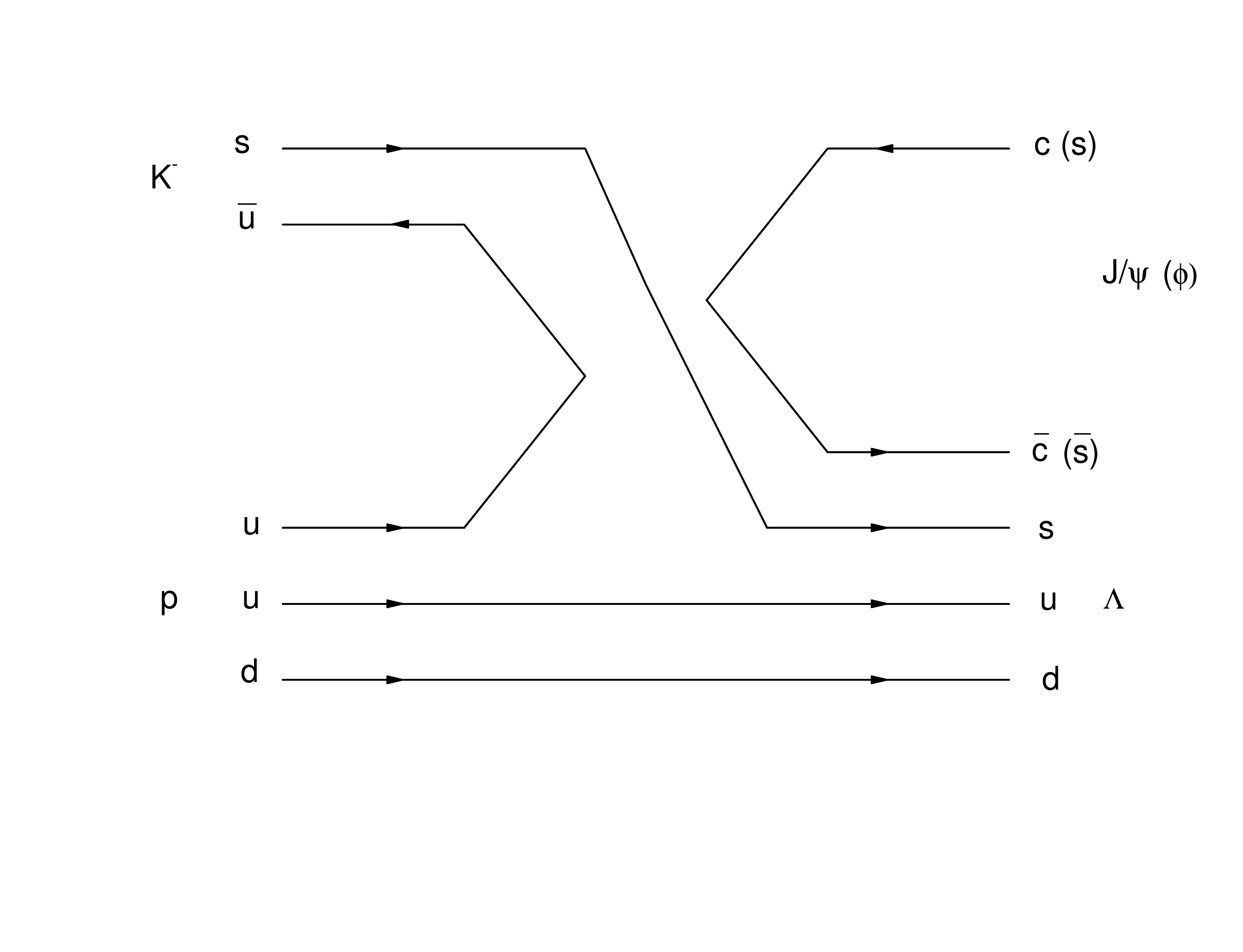}
\vspace*{-2mm} \caption{(Color online) Diagrammatic representation of the $K^-p \to {J/\psi}\Lambda$
and $K^-p \to {\phi}\Lambda$ reactions in terms of the quark lines.}
\label{void}
\end{center}
\end{figure}
The parameter $a$ in the formula (14) can be determined from the normalization condition
\begin{equation}
\int\limits_{4\pi}^{}a{\rm e}^{b_{J/\psi}(t-t^+)}d{\bf \Omega}_{J/\psi}^*=1.
\end{equation}
This, with accounting for Eq. (20), yields
\begin{equation}
a=\frac{p^*_{K^-}p^*_{J/\psi}b_{J/\psi}}{\pi}\left[1-{\rm e}^{-4p^*_{K^-}p^*_{J/\psi}b_{J/\psi}}\right]^{-1}.
\end{equation}

      We now focus on the "in-medium" total cross section
$\sigma_{{K^-}p\to {J/\psi}\Lambda}(\sqrt{s^*})$ for $J/\psi$ production in reaction (1).
According to the aforesaid, it is equivalent to the vacuum cross section
$\sigma_{{K^-}p \to {J/\psi}\Lambda}(\sqrt{s})$, in which the free space center-of-mass energy squared s,
presented by the formula (9), is replaced by the in-medium expression (7). For the free total cross section
$\sigma_{{K^-}p \to {J/\psi}\Lambda}(\sqrt{s})$ no data are also available, so we have to rely on some
indirect way to estimate it. In particular, as a first step on this way, one can make an estimate of the
$J/\psi/\phi$ inclusive total cross section ratio, based on the existing only one experimental data point
(4.4$\pm$3.0) nb for the total cross section $\sigma_{K^-p \to {J/\psi}X}$
of the process $K^-p \to {J/\psi}X$ and on the two data points
(0.69414$\pm$0.15259) mb and (0.65$\pm$0.10) mb for the total cross section $\sigma_{K^-p \to {\phi}X}$
of the reaction $K^-p \to {\phi}X$, measured, respectively, at the closest to each other high excess
energies $\epsilon_{{J/\psi}\Lambda}=\sqrt{s}-\sqrt{s_{\rm th}}=4.462$ GeV and
$\epsilon_{{\phi}\Lambda}=\sqrt{{\tilde s}}-\sqrt{{\tilde s}_{\rm th}}=5.687$ GeV above the
${J/\psi}\Lambda$ and ${\phi}\Lambda$ production thresholds $\sqrt{s_{\rm th}}$ and
$\sqrt{{\tilde s}_{\rm th}}=m_{\phi}+m_{\Lambda}=2.135$ GeV [56] (or at the beam momenta of 39.5 and 32 GeV/c,
correspondingly). Thus, at these excess energies we have
\begin{equation}
 \sigma_{{K^-}p \to {J/\psi}X}/
 \sigma_{{K^-}p \to {\phi}X} \approx 7\cdot10^{-6}.
\end{equation}
Then, accounting for the commonality in the $J/\psi$ and $\phi$ production in ${K^-}p$ interactions,
we assume that the considered ratio of the total cross sections of the inclusive reactions ${K^-}p \to {J/\psi}X$
and ${K^-}p \to {\phi}X$ determined, respectively, at the same excess energies $\epsilon_{{J/\psi}\Lambda}$ and $\epsilon_{{\phi}\Lambda}$ in the threshold regions $\epsilon_{{J/\psi}\Lambda}$ $\le$ 0.4 GeV and
$\epsilon_{{\phi}\Lambda}$ $\le$ 0.4 GeV of our interest
is the same as that of Eq. (28), derived at the high excess energies (cf. Ref. [60]).
On the other hand, in line with Ref. [61] we suppose that this near-threshold
$J/\psi/\phi$ inclusive total cross section ratio is the same as that of the total cross sections of the exclusive reactions ${K^-}p \to {J/\psi}\Lambda$ and ${K^-}p \to {\phi}\Lambda$ also determined, respectively, at the same excess energies $\epsilon_{{J/\psi}\Lambda}$ and $\epsilon_{{\phi}\Lambda}$ in these threshold regions, i.e.
\begin{equation}
 \sigma_{{K^-}p \to {J/\psi}X}(\sqrt{s})/
 \sigma_{{K^-}p \to {\phi}X}(\sqrt{{\tilde s}})=
 \sigma_{{K^-}p \to {J/\psi}\Lambda}(\sqrt{s})/
 \sigma_{{K^-}p \to {\phi}\Lambda}(\sqrt{{\tilde s}}) \approx 7\cdot10^{-6},
\end{equation}
where, according to the preceding discussion, the center-of-mass total energies $\sqrt{s}$ and $\sqrt{{\tilde s}}$
are linked by the relation:
\begin{equation}
\epsilon_{{\phi}\Lambda}=\sqrt{{\tilde s}}-\sqrt{{\tilde s}_{\rm th}}=
\epsilon_{{J/\psi}\Lambda}=\sqrt{s}-\sqrt{s_{\rm th}}.
\end{equation}
With this, we have
\begin{equation}
\sqrt{{\tilde s}}=\sqrt{s}-\sqrt{s_{\rm th}}+\sqrt{{\tilde s}_{\rm th}}=\sqrt{s}-m_{J/\psi}+m_{\phi}.
\end{equation}
At incident antikaon
energies $\sqrt{s} \le 4.6126$ GeV of interest, the c.m.s. energy $\sqrt{{\tilde s}} \le 2.535$ GeV.
The latter one corresponds, as is easy to see, to the laboratory $K^-$ momenta in the region
${\tilde p}_{K^-}$ $\le$ 2.782 GeV/c.
For the free total cross section $\sigma_{{K^-}p \to {\phi}\Lambda}({\sqrt{{\tilde s}}})$
in this region we have adopted the following parametrization of the available here experimental
information [62] on it:
\begin{equation}
\sigma_{{K^-}p \to {\phi}\Lambda}({\sqrt{{\tilde s}}})=Bp^*_{\phi}{\rm e}^{-{\beta}p^*_{\phi}},
\end{equation}
suggested in Ref. [57] (Model 2).
Here, the $\phi$ c.m. momentum $p^*_{\phi}$ is defined above by Eq. (25) and parameters $B$ and $\beta$
are: $B=315.31$ $\mu$b/(GeV/c), $\beta=1.45$ (GeV/c)$^{-1}$.
\begin{figure}[htb]
\begin{center}
\includegraphics[width=16.0cm]{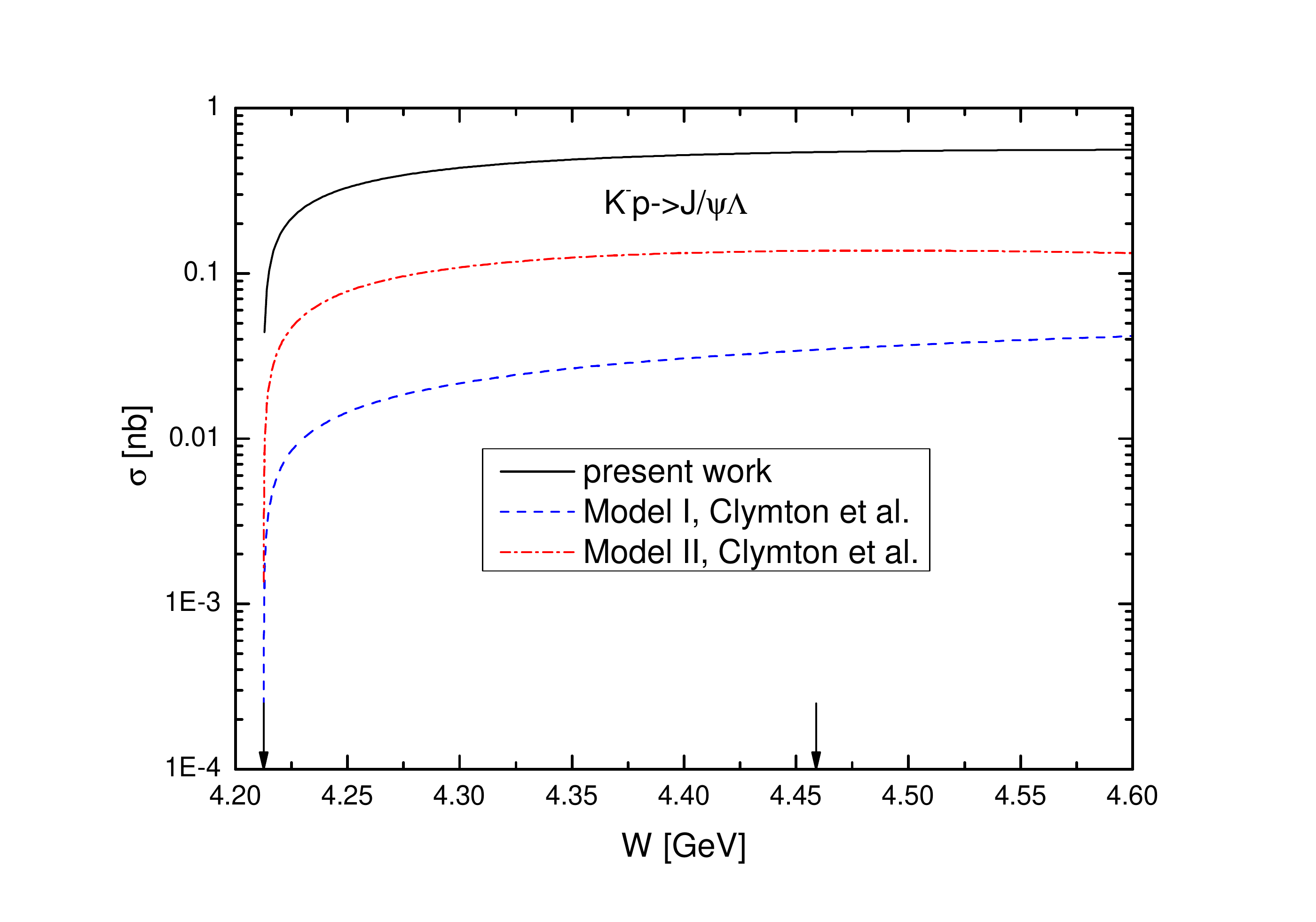}
\vspace*{-2mm} \caption{(Color online) The non-resonant total cross section for the reaction
${K^-}p \to {J/\psi}\Lambda$ as a function of the center-of-mass energy $W=\sqrt{s}$
of the antikaon--proton collisions.
Solid and dashed, dotted-dashed curves are calculations by (29)--(32) and
within the Models I, II developed in Ref. [43], respectively.
The left and right arrows indicate, correspondingly, the center-of-mass threshold energy of 4.2126 GeV
for direct $J/\psi$ production on a free target proton being at rest
and the resonant energy of 4.4588 GeV.}
\label{void}
\end{center}
\end{figure}
\begin{figure}[htb]
\begin{center}
\includegraphics[width=16.0cm]{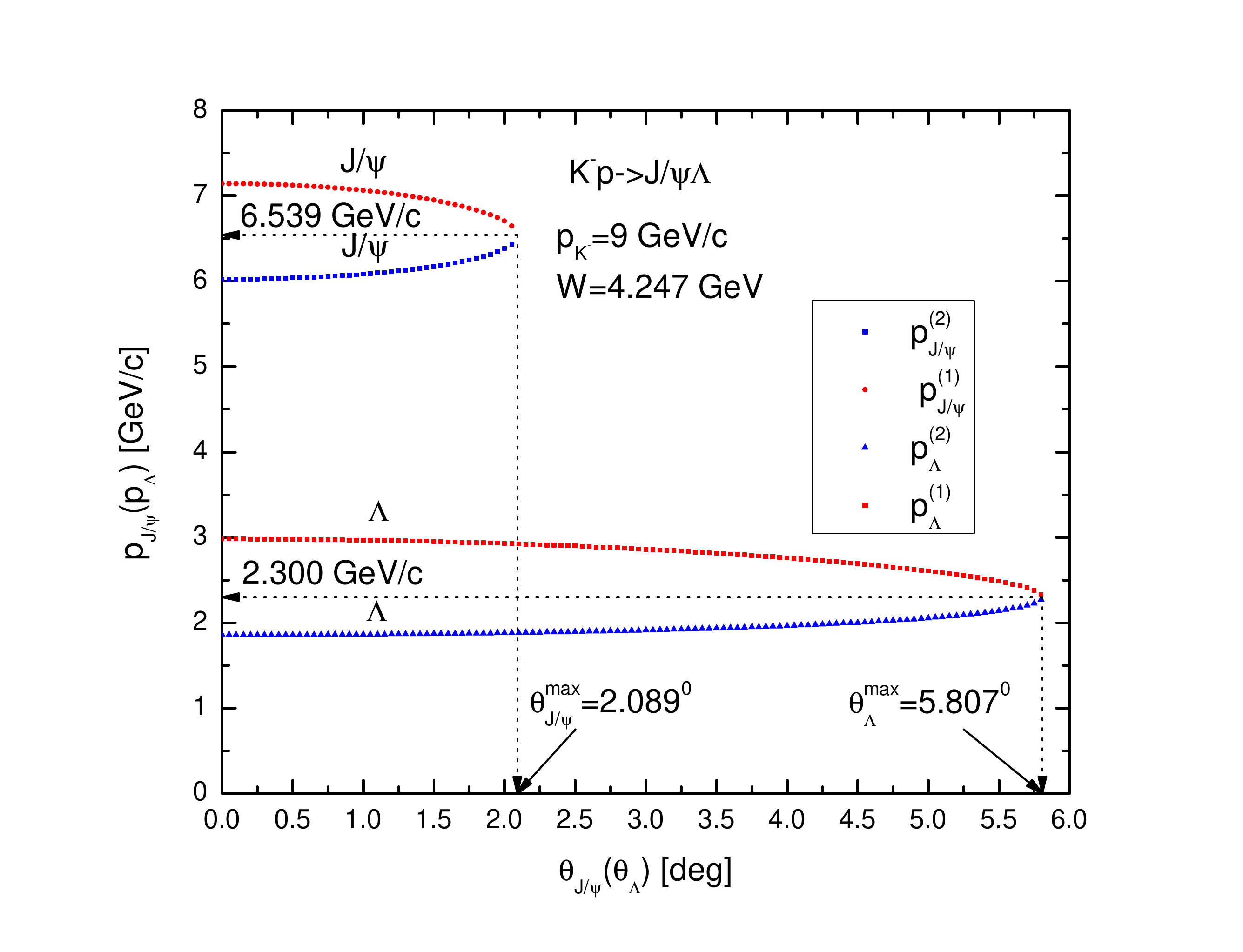}
\vspace*{-2mm} \caption{(Color online) The kinematically allowed $J/\psi$ and $\Lambda$ momenta in the free space non-resonant ${K^-}p \to {J/\psi}\Lambda$ reaction in the laboratory system
at incident antikaon momentum of 9 GeV/c against their exit angles with respect to the beam direction in this system. Left and right vertical arrows mark the maximum values of these angles allowed in the reaction at given antikaon momentum. Upper and lower horizontal arrows indicate the values of the $J/\psi$ meson and $\Lambda$ hyperon momenta,
corresponding to these maximum values.}
\label{void}
\end{center}
\end{figure}
\begin{figure}[htb]
\begin{center}
\includegraphics[width=16.0cm]{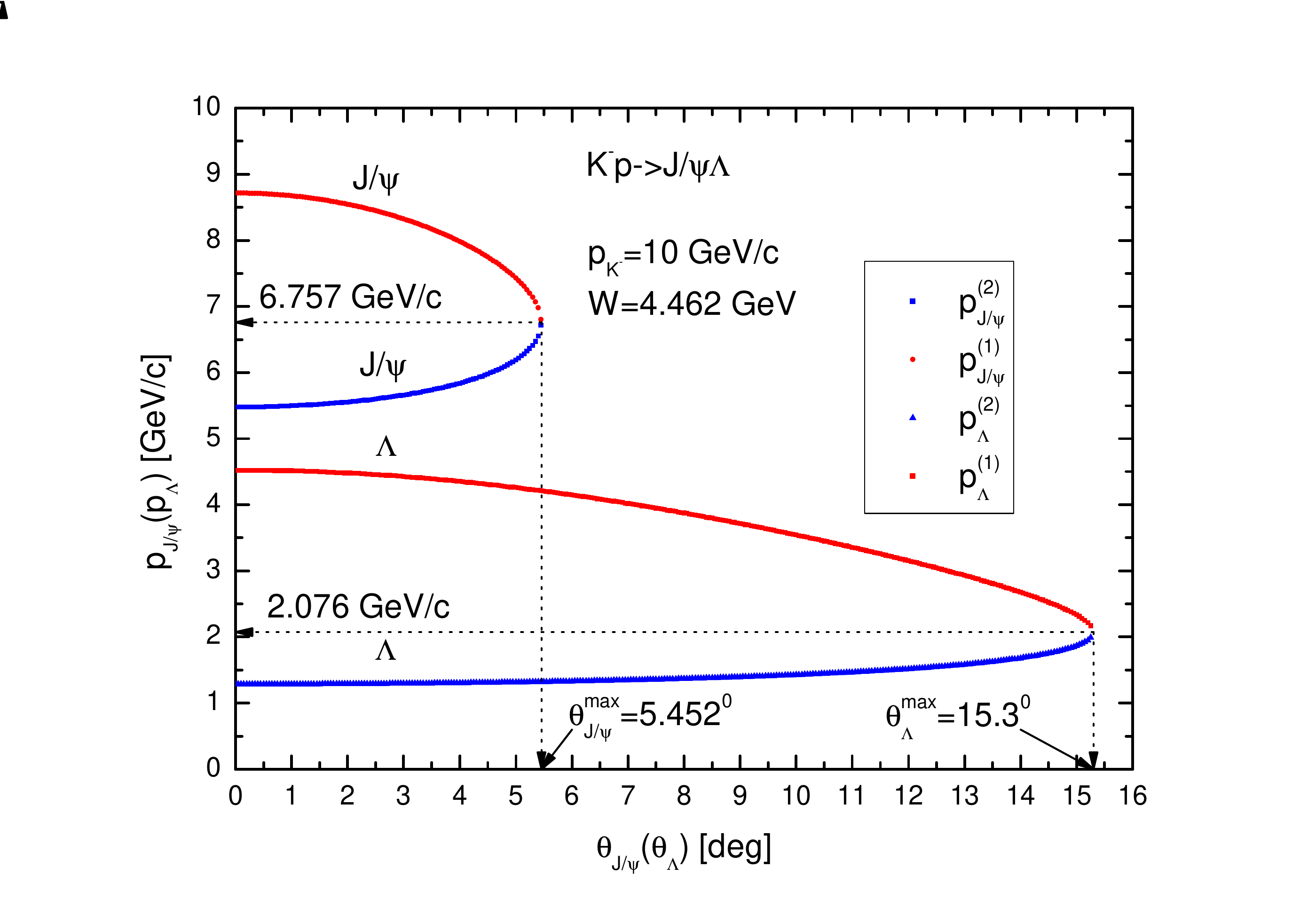}
\vspace*{-2mm} \caption{(Color online) The same as in Fig. 3, but for the initial antikaon momentum of 10 GeV/c.}
\label{void}
\end{center}
\end{figure}

   The results of calculations by Eqs. (29)--(32) of the non-resonant total cross section of the reaction
${K^-}p \to {J/\psi}\Lambda$ at the considered c.m. total energies are shown in Fig. 2 (solid curve).
In this figure we also show the predictions from the effective Lagrangian (Model I) and Regge (Model II) approaches [43]
(dashed and dotted-dashed curves, respectively)
\footnote{$^)$The author thanks S. Clymton for sending these predictions to him.}$^)$
.
We see that while all the results, presented in Fig. 2, show a similar behavior, they differ significantly in
strength among themselves. Thus, at antikaon c.m. energies in the vicinity of the resonant energy $W=4.4588$ GeV our parametrization (29)--(32) predicts an experimentally measurable background total cross section of the order of about 1 nb and it is considerably larger (by factor of approximately 16) than the results from the effective Lagrangian method (Model I) [43]. Whereas our calculations are much closer here to the results from the another sophisticated theoretical hadronic approach developed in Ref. [43] - Regge approach (Model II), namely: they are larger here than the latter ones only by a factor of about 4. This enables us to hope that the use of the  phenomenological model, described above, for estimating the elastic background under the $P_{cs}(4459)^0$ pentaquark peak is correct and meaningful. And these estimates may serve as guidance for future dedicated experiment at the J-PARC.

As was noted above (see also below), at the considered incident antikaon momenta the
$J/\psi$ mesons are produced at small angles with respect to the $K^-$ beam direction.
Therefore, we will calculate the $J/\psi$ momentum
differential distributions from $^{12}$C and $^{184}$W targets
for the laboratory solid angle
${\Delta}{\bf \Omega}_{J/\psi}$=$0^{\circ} \le \theta_{J/\psi} \le 20^{\circ}$,
and $0 \le \varphi_{J/\psi} \le 2{\pi}$.
Here, $\varphi_{J/\psi}$ is the azimuthal angle of the $J/\psi$ momentum ${\bf p}_{J/\psi}$
in the laboratory system with $z$-axis directed along the momentum ${\bf p}_{K^-}$ of the incoming
antikaon beam.
Then, integrating the full inclusive differential cross section (2) over this angular domain,
we can represent the differential cross section for $J/\psi$ meson
production in antikaon-induced reactions from the direct process (1),
corresponding to this angular domain, in the following form:
\begin{equation}
\frac{d\sigma_{{K^-}A\to {J/\psi}X}^{({\rm dir})}
(p_{K^-},p_{J/\psi})}{dp_{J/\psi}}=
\int\limits_{{\Delta}{\bf \Omega}_{J/\psi}}^{}d{\bf \Omega}_{J/\psi}
\frac{d\sigma_{{K^-}A\to {J/\psi}X}^{({\rm dir})}
({\bf p}_{K^-},{\bf p}_{J/\psi})}{d{\bf p}_{J/\psi}}p_{J/\psi}^2
\end{equation}
$$
=2{\pi}\left(\frac{Z}{A}\right)I_{V}[A,\sigma_{{J/\psi}N}]
\int\limits_{\cos20^{\circ}}^{1}d\cos{{\theta_{J/\psi}}}
\left<\frac{d\sigma_{{K^-}p\to {J/\psi}{\Lambda}}(p_{K^-},
p_{J/\psi},\theta_{J/\psi})}{dp_{J/\psi}d{\bf \Omega}_{J/\psi}}\right>_A.
$$

  Before closing this subsection, let us more closely consider, using the relativistic kinematics,
more simpler case of the production of $J/\psi$ mesons and $\Lambda$ hyperons
in the elementary reaction ${K^-}p \to {J/\psi}\Lambda$ proceeding on a free target proton being at rest
to get some feeling about their kinematic characteristics allowed in this reaction at incident
antikaon momenta of interest. As known from the kinematics of two-body reaction with a threshold
(as in present our case), the laboratory polar $J/\psi$ and $\Lambda$ production angles $\theta_{J/\psi}$
and $\theta_{\Lambda}$ range from 0 to a maximal values $\theta^{\rm max}_{J/\psi}$ and
$\theta^{\rm max}_{\Lambda}$, respectively, i.e.:
\begin{equation}
     0 \le \theta_{J/\psi} \le \theta^{\rm max}_{J/\psi},
\end{equation}
\begin{equation}
     0 \le \theta_{\Lambda} \le \theta^{\rm max}_{\Lambda};
\end{equation}
where
\begin{equation}
 \theta^{\rm max}_{J/\psi}={\rm arcsin}[(\sqrt{s}p^{*}_{J/\psi})/(m_{J/\psi}p_{K^-})],
\end{equation}
\begin{equation}
 \theta^{\rm max}_{\Lambda}={\rm arcsin}[(\sqrt{s}p^{*}_{\Lambda})/(m_{\Lambda}p_{K^-})].
\end{equation}
Here, the $J/\psi$ c.m. momentum $p^*_{J/\psi}$ is defined above by Eq. (18),
in which the in-medium center-of-mass energy squared $s^*$ should be replaced by the
free space squared invariant collision energy $s$ determined by the formula (9); analogously,
$p^*_{\Lambda}$ is the $\Lambda$ c.m. momentum, which is equal to the $J/\psi$ c.m. momentum $p^*_{J/\psi}$.
Eqs. (34)--(37) imply that the following inequalities are fulfilled:
\begin{equation}
(\sqrt{s}p^{*}_{J/\psi})/(m_{J/\psi}p_{K^-}) \le 1,\,\,\,\,
 (\sqrt{s}p^{*}_{\Lambda})/(m_{\Lambda}p_{K^-}) \le 1.
\end{equation}
Taking into consideration the relativistic relations between the $J/\psi$, $\Lambda$ momenta $p^*_{J/\psi}$,
$p^*_{\Lambda}$ in the $K^-p$ c.m.s. and their velocities $v^{*}_{J/\psi}$, $v^{*}_{\Lambda}$ in this system
\begin{equation}
p^{*}_{J/\psi}=m_{J/\psi}v^{*}_{J/\psi}/\sqrt{1-(v^{*}_{J/\psi})^2},\,\,\,\,
p^{*}_{\Lambda}=m_{\Lambda}v^{*}_{\Lambda}/\sqrt{1-(v^{*}_{\Lambda})^2},
\end{equation}
these inequalities can be easily transformed to the following equivalent ones between the
velocity $v_{\rm cm}$ of the ${K^-}p$ center-of-mass system in the laboratory frame and the above
velocities:
\begin{equation}
v_{\rm cm} \ge v^{*}_{J/\psi},\,\,\,\,\,v_{\rm cm} \ge v^{*}_{\Lambda},
\end{equation}
where
\begin{equation}
v_{\rm cm}=p_{K^-}/(E_{K^-}+m_p),\,\,\,\,
v^{*}_{J/\psi}=p^{*}_{J/\psi}/E^{*}_{J/\psi},\,\,\,\,
v^{*}_{\Lambda}=p^{*}_{\Lambda}/E^{*}_{\Lambda},\,\,\,\,
E^{*}_{\Lambda}=\sqrt{m^2_{\Lambda}+p^{*2}_{\Lambda}}
\end{equation}
and the $J/\psi$ total c.m. energy $E^{*}_{J/\psi}$ is determined above by Eq. (16).
From Eqs. (36), (37), (41) we get, for instance, that $\theta^{\rm max}_{J/\psi}=2.089^{\circ}$,
$\theta^{\rm max}_{\Lambda}=5.807^{\circ}$,
$v_{\rm cm}=0.904$, $v^{*}_{J/\psi}=0.077$, $v^{*}_{\Lambda}=0.210$
at incident $K^-$ beam momentum of $p_{K^-}=9$ GeV/c (or at the excess energy
$\sqrt{s}-\sqrt{s_{{\rm th}}}=0.035$ GeV) and
$\theta^{\rm max}_{J/\psi}=5.452^{\circ}$, $\theta^{\rm max}_{\Lambda}=15.3^{\circ}$,
$v_{\rm cm}=0.913$, $v^{*}_{J/\psi}=0.208$, $v^{*}_{\Lambda}=0.509$ at $K^-$
momentum of $p_{K^-}=10$ GeV/c (at which $\sqrt{s}-\sqrt{s_{{\rm th}}}=0.250$ GeV).
So, as we approach threshold momentum (8.844 GeV/c), the $J/\psi$'s and
$\Lambda$'s become increasingly focused in a forward cones along the beam direction in the laboratory frame.
And, indeed, the velocity $v_{\rm cm}$ of the $K^-p$ c.m.s. in the laboratory frame
is greater than the $J/\psi$ meson and the $\Lambda$ hyperon c.m. velocities $v^{*}_{J/\psi}$
and $v^{*}_{\Lambda}$ at these initial antikaon momenta. Imposing energy-momentum conservation in the
free space process (1), we find two different solutions for the laboratory $J/\psi$ meson and $\Lambda$
hyperon momenta $p_{J/\psi}$ and $p_{\Lambda}$ at given production angles $\theta_{J/\psi}$ and
$\theta_{\Lambda}$, belonging, respectively, to the angular intervals (34) and (35):
\begin{equation}
p^{(1,2)}_{J/\psi}(\theta_{J/\psi})=
\frac{p_{K^-}\sqrt{s}E^{*}_{J/\psi}\cos{\theta_{J/\psi}}\pm(E_{K^-}+m_p)\sqrt{s}\sqrt{p^{*2}_{J/\psi}-{\gamma^2_{\rm cm}}{v^2_{\rm cm}}m^2_{J/\psi}\sin^2{\theta_{J/\psi}}}}{(E_{K^-}+m_p)^2-p^2_{K^-}\cos^2{\theta_{J/\psi}}},
\end{equation}
\begin{equation}
p^{(1,2)}_{\Lambda}(\theta_{\Lambda})=
\frac{p_{K^-}\sqrt{s}E^{*}_{\Lambda}\cos{\theta_{\Lambda}}\pm(E_{K^-}+m_p)\sqrt{s}\sqrt{p^{*2}_{\Lambda}-{\gamma^2_{\rm cm}}{v^2_{\rm cm}}m^2_{\Lambda}\sin^2{\theta_{\Lambda}}}}{(E_{K^-}+m_p)^2-p^2_{K^-}\cos^2{\theta_{\Lambda}}}.
\end{equation}
Here, ${\gamma_{\rm cm}}=(E_{K^-}+m_p)/\sqrt{s}$, sign "+" in the numerators of Eqs. (42), (43)
corresponds to the first solutions $p^{(1)}_{J/\psi}$, $p^{(1)}_{\Lambda}$
and sign "-" - to the second ones $p^{(2)}_{J/\psi}$, $p^{(2)}_{\Lambda}$.
In the case of zero $J/\psi$ and $\Lambda$ production angles the expressions (42), (43) are reduced
to the following more simpler forms:
\begin{equation}
p^{(1,2)}_{J/\psi}(0^{\circ})=
{\gamma_{\rm cm}}E^{*}_{J/\psi}(v_{\rm cm}\pm v^{*}_{J/\psi}),
\end{equation}
\begin{equation}
p^{(1,2)}_{\Lambda}(0^{\circ})=
{\gamma_{\rm cm}}E^{*}_{\Lambda}(v_{\rm cm}\pm v^{*}_{\Lambda}).
\end{equation}
As is easy to see, the momenta (44), (45) correspond to the respective $J/\psi$ and $\Lambda$ total energies:
\begin{equation}
E^{(1,2)}_{J/\psi}(0^{\circ})=
{\gamma_{\rm cm}}[E^{*}_{J/\psi} \pm v_{\rm cm}p^{*}_{J/\psi}],
\end{equation}
\begin{equation}
E^{(1,2)}_{\Lambda}(0^{\circ})=
{\gamma_{\rm cm}}[E^{*}_{\Lambda} \pm v_{\rm cm}p^{*}_{\Lambda}].
\end{equation}
Interestingly, that
\begin{equation}
p^{(1)}_{J/\psi}(0^{\circ})+p^{(2)}_{\Lambda}(0^{\circ})=p_{K^-},\,\,\,\,
E^{(1)}_{J/\psi}(0^{\circ})+E^{(2)}_{\Lambda}(0^{\circ})=E_{K^-}+m_p
\end{equation}
and
\begin{equation}
p^{(2)}_{J/\psi}(0^{\circ})+p^{(1)}_{\Lambda}(0^{\circ})=p_{K^-},\,\,\,\,
E^{(2)}_{J/\psi}(0^{\circ})+E^{(1)}_{\Lambda}(0^{\circ})=E_{K^-}+m_p.
\end{equation}
As should have been, the relations (48) and (49) express the energy-momentum conservation in the process (1),
taking place on a free target proton at rest, in the laboratory frame for zero $J/\psi$ and $\Lambda$
production angle. They correspond, respectively, to
the $J/\psi$ going forward (at 0$^{\circ}$) and $\Lambda$ going backward (at 180$^{\circ}$)
in the c.m. system and vice versa. Being guided by Eqs. (44)--(47), for $p_{K^-}=9$ GeV/c we have:
\begin{equation}
p^{(1)}_{J/\psi}(0^{\circ})=7.143~{\rm GeV/c},\,\,\,p^{(2)}_{J/\psi}(0^{\circ})=6.022~{\rm GeV/c};
\end{equation}
$$
p^{(1)}_{\Lambda}(0^{\circ})=2.978~{\rm GeV/c},\,\,\, p^{(2)}_{\Lambda}(0^{\circ})=1.857~{\rm GeV/c}
$$
and
\begin{equation}
E^{(1)}_{J/\psi}(0^{\circ})=7.785~{\rm GeV},\,\,\, E^{(2)}_{J/\psi}(0^{\circ})=6.772~{\rm GeV};
\end{equation}
$$
E^{(1)}_{\Lambda}(0^{\circ})=3.180~{\rm GeV},\,\,\, E^{(2)}_{\Lambda}(0^{\circ})=2.167~{\rm GeV}.
$$
For $p_{K^-}=10$ GeV/c we also get:
\begin{equation}
p^{(1)}_{J/\psi}(0^{\circ})=8.714~{\rm GeV/c},\,\,\,p^{(2)}_{J/\psi}(0^{\circ})=5.478~{\rm GeV/c};
\end{equation}
$$
p^{(1)}_{\Lambda}(0^{\circ})=4.522~{\rm GeV/c},\,\,\, p^{(2)}_{\Lambda}(0^{\circ})=1.286~{\rm GeV/c}
$$
and
\begin{equation}
E^{(1)}_{J/\psi}(0^{\circ})=9.248~{\rm GeV},\,\,\, E^{(2)}_{J/\psi}(0^{\circ})=6.292~{\rm GeV};
\end{equation}
$$
E^{(1)}_{\Lambda}(0^{\circ})=4.658~{\rm GeV},\,\,\, E^{(2)}_{\Lambda}(0^{\circ})=1.702~{\rm GeV},
$$
so that the relations (48), (49) are, indeed, satisfied at these initial $K^-$ momenta
\footnote{$^)$At them, the total energy in the entrance channel $E_{K^-}+m_p$ is equal to 9.952 GeV and
to 10.950 GeV, correspondingly.}$^)$
.
An inspection of the expressions (42), (43) tells us that the first solutions $p^{(1)}_{J/\psi}$ and
$p^{(1)}_{\Lambda}$ as well as the second ones $p^{(2)}_{J/\psi}$ and $p^{(2)}_{\Lambda}$
exhibit different dependences, respectively, on the production angles $\theta_{J/\psi}$ and $\theta_{\Lambda}$
within the angular intervals [0,$\theta_{J/\psi}^{\rm max}]$ and [0,$\theta_{\Lambda}^{\rm max}]$.
While the former fall off as the production angles
$\theta_{J/\psi}$ and $\theta_{\Lambda}$ increase here, the latter ones increase as these angles
also increase here (see, Figs. 3 and 4 as well) and
\begin{equation}
p^{(1)}_{J/\psi}(\theta_{J/\psi}^{\rm max})=p^{(2)}_{J/\psi}(\theta_{J/\psi}^{\rm max})=
p_{J/\psi}(\theta_{J/\psi}^{\rm max}),
\end{equation}
$$
p^{(1)}_{\Lambda}(\theta_{\Lambda}^{\rm max})=p^{(2)}_{\Lambda}(\theta_{\Lambda}^{\rm max})=
p_{\Lambda}(\theta_{\Lambda}^{\rm max}),
$$
 where
\begin{equation}
p_{J/\psi}(\theta_{J/\psi}^{\rm max})=(p_{K^-}m_{J/\psi}^2\cos{\theta_{J/\psi}^{\rm max}})/
(\sqrt{s}E^{*}_{J/\psi}),\,\,\,
p_{\Lambda}(\theta_{\Lambda}^{\rm max})=(p_{K^-}m_{\Lambda}^2\cos{\theta_{\Lambda}^{\rm max}})/
(\sqrt{s}E^{*}_{\Lambda}).
\end{equation}
With these, the $J/\psi$ meson and $\Lambda$ hyperon total energies $E_{J\psi}(\theta^{\rm max}_{J/\psi})$
and $E_{\Lambda}(\theta^{\rm max}_{\Lambda})$ in the l.s.,
corresponding to their maximal production angles $\theta^{\rm max}_{J/\psi}$ and $\theta^{\rm max}_{\Lambda}$,
can be represented in the following simple forms:
\begin{equation}
E_{J/\psi}(\theta^{\rm max}_{J/\psi})={\gamma_{\rm cm}}m^2_{J/\psi}/E^{*}_{J/\psi},\,\,\,\,
E_{\Lambda}(\theta^{\rm max}_{\Lambda})={\gamma_{\rm cm}}m^2_{\Lambda}/E^{*}_{\Lambda}.
\end{equation}
In line with Eqs. (55) and (56), for $p_{K^-}=9$ GeV/c we obtain then that
$p_{J/\psi}(\theta_{J/\psi}^{\rm max})=6.539$ GeV/c, $E_{J/\psi}(\theta^{\rm max}_{J/\psi})=7.235$ GeV and
$p_{\Lambda}(\theta_{\Lambda}^{\rm max})=2.300$ GeV/c, $E_{\Lambda}(\theta^{\rm max}_{\Lambda})=2.556$ GeV.
For $p_{K^-}=10$ GeV/c we have
$p_{J/\psi}(\theta_{J/\psi}^{\rm max})=6.757$ GeV/c, $E_{J/\psi}(\theta^{\rm max}_{J/\psi})=7.433$ GeV and
$p_{\Lambda}(\theta_{\Lambda}^{\rm max})=2.076$ GeV/c, $E_{\Lambda}(\theta^{\rm max}_{\Lambda})=2.357$ GeV
(cf. Figs. 3 and 4).

 Taking into account the above considerations and the results, presented in Figs. 3, 4, we can conclude
that, for example, the kinematically allowed $J/\psi$ meson laboratory momenta and total energies in the free space
non-resonant ${K^-}p \to {J/\psi}\Lambda$  reaction at given incident antikaon momentum
vary within the following momentum and energy ranges:
\begin{equation}
p^{(2)}_{J/\psi}(0^{\circ}) \le p_{J/\psi} \le p^{(1)}_{J/\psi}(0^{\circ}),
\end{equation}
\begin{equation}
E^{(2)}_{J/\psi}(0^{\circ}) \le E_{J/\psi} \le E^{(1)}_{J/\psi}(0^{\circ}),
\end{equation}
where the quantities $p^{(1,2)}_{J/\psi}(0^{\circ})$ and $E^{(1,2)}_{J/\psi}(0^{\circ})$ are defined above
by Eqs. (44) and (46), correspondingly. In analogy to $J/\psi$ production, for $\Lambda$ hyperon, the kinematically allowed momentum and energy intervals look like those of (57) and (58), but in which one needs to make the substitution $J/\psi$ $\to$ $\Lambda$.

  It is of further interest to calculate the $J/\psi$ energy spectrum
$d\sigma_{{K^-}p \to {J/\psi}{\Lambda}}[\sqrt{s},p_{J/\psi}]/ dE_{J/\psi}$
from the considered reaction ${K^-}p \to {J/\psi}\Lambda$ as a function of
the $J/\psi$ total energy $E_{J/\psi}$ belonging to the interval (58).
The differential cross section
$d\sigma_{{K^-}p \to {J/\psi}{\Lambda}}[\sqrt{s},{\bf p}_{J/\psi}]/ d{\bf p}_{J/\psi}$
of this reaction can be obtained from more general one (10) in the limits: ${\bf p}_t \to 0$,
$E_t \to m_p$ and $s^* \to s$. The integration of this cross section over the angle $\theta_{J/\psi}$
between the momenta ${\bf p}_{K^-}$ and ${\bf p}_{J/\psi}$ when this angle lies in the allowed angular
region (34) with accounting for the properties of the energy conserving Dirac $\delta$-function and the
relation $d{\bf p}_{J/\psi}=p_{J/\psi}E_{J/\psi}dE_{J/\psi}d\Omega_{J/\psi}$ yields:
\begin{equation}
\frac{d\sigma_{{K^-}p \to {J/\psi}{\Lambda}}[\sqrt{s},p_{J/\psi}]}{dE_{J/\psi}}=
2\pi
\int\limits_{\cos{\theta_{J/\psi}^{\rm max}}}^{1}d\cos{\theta_{J/\psi}}p_{J/\psi}E_{J/\psi}
\frac{d\sigma_{{K^-}p\to {J/\psi}\Lambda}[\sqrt{s},{\bf p}_{J/\psi}]}{d{\bf p}_{J/\psi}}=
\end{equation}
$$
=
\left(\frac{2{\pi}\sqrt{s}}{p_{K^-}p^{*}_{J/\psi}}\right)
\frac{d\sigma_{{K^{-}}p \to {J/\psi}{\Lambda}}[\sqrt{s},\theta_{J/\psi}^*(x_0)]}{d{\bf \Omega}_{J/\psi}^*}~{\rm for}
~E^{(2)}_{J/\psi}(0^{\circ}) \le E_{J/\psi} \le E^{(1)}_{J/\psi}(0^{\circ}),
$$
where
\begin{equation}
x_0=\frac{[p^2_{K^-}+p^2_{J/\psi}+m^2_{\Lambda}-(\omega+m_p)^2]}{2p_{K^-}p_{J/\psi}},\,\,\,\,\,
p_{J/\psi}=\sqrt{E^2_{J/\psi}-m^2_{J/\psi}}
\end{equation}
and the quantity $\cos{\theta_{J/\psi}^*(x_0)}$ is determined by Eq. (21), in which one has to make the
substitution: $\cos{\theta_{J/\psi}} \to x_0$, and the $K^-$ and $J/\psi$ c.m. momenta $p^*_{K^-}$
and $p^*_{J/\psi}$ are defined by formulas (17) and (18), respectively, in which, in-line with
the above-mentioned, one needs also to perform the replacements: $E_t \to m_p$, $p_t \to 0$ and $s^* \to s$.
The expression (59) will be used by us for evaluating the $J/\psi$ energy spectrum, arising from the reaction
$K^-p \to {J/\psi}\Lambda$ proceeding on a free space proton at rest, for incident beam momenta of 9 and
10 GeV/c (see below).

\subsection*{2.2. Two-step resonant $J/\psi$ production mechanism}

  At $K^-$ meson center-of-mass excess energies $\sqrt{s}-\sqrt{s_{\rm th}}$ $\le$ 0.4 GeV
of our interest, an incident antikaon can produce a
neutral hidden-charm strange pentaquark $P_{cs}(4459)^0$ with pole mass $M_{cs}=4458.8$ MeV
directly in the first inelastic collision with an intranuclear proton
\footnote{$^)$We remind that the threshold (resonant) momentum $p^{\rm R}_{K^-}$
for the production of $P_{cs}(4459)^0$ pentaquark on a free target proton being at rest
is $p^{\rm R}_{K^-}=9.983$ GeV/c.}$^)$
:
\begin{equation}
{K^-}+p \to P_{cs}(4459)^0.
\end{equation}
Then the produced intermediate hidden-charm pentaquark resonance can decay into the final state
$J/\psi$$\Lambda$:
\begin{equation}
P_{cs}(4459)^0 \to J/\psi+\Lambda.
\end{equation}
Presently, as was noted above, neither the branching ratio $Br[P_{cs}(4459)^0 \to {J/\psi}\Lambda]$ of this decay,
nor spin-parity quantum numbers of the $P_{cs}(4459)^0$ state with strangeness $S=-1$
have been determined yet experimentally because of a limited statistics in the experiment [24].
Therefore, to evaluate the $J/\psi$ production cross sections from the production/decay sequence (61), (62),
taking place both on a vacuum proton and on a proton embedded in a nuclear target,
one must rely on the theoretical predictions. Among them, there are, in particular, the following theoretical
guidelines, based on the fact that the pole mass of the $P_{cs}(4459)^0$ state is just about 19 MeV below the $\Xi^0_c{\bar D}^{*0}$ threshold [24], so it is quite natural to consider it as the $\Xi_c{\bar D}^*$ molecular
state with the spin-parity quantum numbers $J^P=(1/2)^-$ or $(3/2)^-$.
Thus, the recent study [63] of the $P_{cs}(4459)^0$, using the method of QCD sum rules, supports its interpretation
as the $\Xi_c{\bar D}^*$ hadronic molecular state located at 4.46$^{+0.16}_{-0.14}$ GeV for spin-parity
$J^P=(1/2)^-$ and at 4.47$^{+0.19}_{-0.15}$ GeV for $J^P=(3/2)^-$. On the other hand, in Ref. [64],
the $P_{cs}(4459)^0$ resonance was studied in the diquark-diquark-antiquark picture again with the QCD sum rules.
The results obtained here support assigning the $P_{cs}(4459)^0$ to be hidden-charm compact pentaquark state with
the spin-parity $J^P=(1/2)^-$. The QCD sum rule approach was also exploited in Ref. [65] to calculate the magnetic
dipole moments of the $P_c(4440)$, $P_c(4457)$ and $P_{cs}(4459)$ pentaquark states by considering them as the
diquark-diquark-antiquark and molecular pictures with quantum numbers $J^P=(3/2)^-$, $J^P=(1/2)^-$ and
$J^P=(1/2)^-$, respectively. Spectrum of the hidden-charm strange molecular pentaquarks $P_{cs}$ has been
studied in Ref. [66] in chiral effective field theory. The authors found ten molecular pentaquarks $P_{cs}$
in the $\Xi_c{\bar D}^*$, $\Xi_{c}{'}{\bar D}^*$, $\Xi_{c}^{*}{\bar D}^*$ systems and
also predicted, in particular, the masses
of the $\Xi_c{\bar D}^*$ bound state for $J=1/2$ and $J=3/2$ to be 4456.9$^{+3.2}_{-3.3}$ MeV and
4463.0$^{+2.8}_{-3.0}$ MeV, respectively. These both two values are consistent with the experimental mass
of the $P_{cs}(4459)^0$, supporting its interpretation as the $\Xi_c{\bar D}^*$ hadronic molecular state with
the spin-parity of either $J^P=(1/2)^-$ or $(3/2)^-$.
Further, the author in Ref. [67], interpreting the pentaquark $P_{cs}(4459)^0$
observed by the LHCb Collaboration [24] as the coupled strange hidden-charm hadronic molecule with the dominant
$s$-wave channels $\Xi_c{\bar D}^*$ and $\Xi_c^*{\bar D}$ with the spin-parity quantum numbers
$J^P=(3/2)^-$, predicted adopting the one-boson-exchange model that
the partial decay width for the process (62) is around 0.1 MeV in the resonance region.
With this value as well as with the $P_{cs}(4459)^0$ resonance total decay width in its rest frame of about
20 MeV also predicted in [67] in this region, we get that the branching fraction
$Br[P_{cs}(4459)^0 \to {J/\psi}\Lambda]$ amounts to 0.5\%. On the other hand, reexamining the results of Ref. [68]
and using the coupled channel unitary approach combined with heavy quark spin symmetry, Ref. [69] also
shows that this resonance can be assigned as a hadronic molecule composed of $\Xi_c{\bar D}^*$ with possible
$J^P$ quantum numbers $(1/2)^-$ or $(3/2)^-$ and the magnitude of the branching fraction of its decay to
the ${J/\psi}\Lambda$ channel is 3.31\% for $J^P=(1/2)^-$ and is 14.68\% for $J^P=(3/2)^-$.
Along the way of $s$-wave $\Xi_c{\bar D}^*$ molecular scenario for the $P_{cs}(4459)^0$ strange pentaquark,
the production mechanism of the $\Xi^-_b \to P_{cs}(4459)^0{K^-}$ decay has been investigated in Ref. [70] adopting
an effective Lagrangian approach and the value of branching ratio of this decay was estimated to be of the order
of 10$^{-4}$. With this value as well as with an experimental estimation of the branching fraction of the
$\Xi^-_b \to P_{cs}(4459)^0{K^-} \to {J/\psi}{\Lambda}K^-$ decay, the branching ratio
$Br[P_{cs}(4459)^0 \to {J/\psi}\Lambda]$ is evaluated in [70] to be of the order of 1 $\sim$ 10\%, which is
consistent with the results of the analysis [71] of the decay properties of the $\Xi_c{\bar D}^*$ bound state
in the ${J/\psi}\Lambda$ mode. By exploiting an effective Lagrangian for the $P_c{J/\psi}$ coupling in combination
with the branching fraction $Br[P_{c}^+ \to {J/\psi}p]$ upper limit of 5\% and with the predicted masses of the predicted pentaquark states, the width of the decay of the compact hidden-charm strange pentaquark denoted as $P_c^{1'0}$
with $|udsc{\bar c}>$ valence quark content
to the ${J/\psi}\Lambda$ channel was computed in Ref. [72]. It turned out to be equal to 8.35 MeV.
Employing this value and the total decay width of
17.3 MeV of the $P_{cs}(4459)^0$ resonance measured by the LHCb Collaboration, the branching ratio
$Br[P_{cs}(4459)^0 \to {J/\psi}\Lambda]$ could be estimated at the level of 50\%. It is also worth mentioning that
the branching ratios of 1, 10 and 50\% of the $P_{cs}(4459)^0 \to {J/\psi}\Lambda$ decay were used in Ref. [43] to
determine the contribution from this decay to the total cross section of the $K^-p \to {J/\psi}\Lambda$ reaction
within the effective Lagrangian method and the Regge approach.
Investigation of the $P_{cs}(4459)^0$ pentaquark via its strong decay to ${J/\psi}\Lambda$ has been performed
in Ref. [73] within the QCD sum rule framework. The partial width of this decay
$\Gamma_{P_{cs}(4459)^0 \to {J/\psi}\Lambda}$ was obtained to be
$\Gamma_{P_{cs}(4459)^0 \to {J/\psi}\Lambda}=(15.87\pm3.11)$ MeV assigning the spin-parity numbers
of $P_{cs}(4459)^0$ state as $J^P=(1/2)^-$ and its structure as diquark-diquark-antiquark. If we adopt
an experimental magnitude of 17.3 MeV for the $P_{cs}(4459)^0$ total decay width $\Gamma_{cs}$,
we obtain with the above
partial decay width that the branching fraction $Br[P_{cs}(4459)^0 \to {J/\psi}\Lambda]$ even could reach
the value of the order of 90\%. Based on the combined effective field theory applied to
the hadronic molecular picture of the $P_{cs}(4459)^0$ pentaquark, Ref. [74] argues
that its spin $J=3/2$ is preferable over $J=1/2$.
Accounting for both this argument and the aforesaid theoretical activities, it is natural
to assign the $P_{cs}(4459)^0$ possible spin-parity quantum numbers $J^P=(3/2)^-$ as well as
to adopt for the branching ratio $Br[P_{cs}(4459)^0 \to {J/\psi}\Lambda]$ of the decay (62)
in our study the following five conservative options: $Br[P_{cs}(4459)^0 \to {J/\psi}\Lambda]=1$, 3, 5, 10, 15
and additional one with enhanced value of 50\% of this ratio
in order to see the size of its impact on the resonant $J/\psi$ yield in ${K^-}p \to {J/\psi}\Lambda$,
${K^-}$$^{12}$C $\to {J/\psi}X$ and ${K^-}$$^{184}$W $\to {J/\psi}X$ reactions.

   Energy-momentum conservation in the process (61), proceeding on a proton embedded in a nuclear target,
leads to the conclusion that the mass of the intermediate $P_{cs}(4459)^0$ resonance is equal to
the total ${K^-}p$ c.m.s. energy $\sqrt{s^*}$ defined above by Eq.~(7).
In line with this, we assume that its in-medium spectral function $S_{cs}(\sqrt{s^*},\Gamma_{cs})$,
is described by the non-relativistic Breit-Wigner distribution (cf. [44, 75--77])
\footnote{$^)$ We neglect, for reasons of numerical simplicity, the medium modification of the
$P_{cs}(4459)^0$ mass and total decay width in the present work.}$^)$
:
\begin{equation}
S_{cs}(\sqrt{s^*},\Gamma_{cs})
=\frac{1}{2\pi}\frac{\Gamma_{cs}}{(\sqrt{s^*}-M_{cs})^2+{\Gamma}_{cs}^2/4}.
\end{equation}
The in-medium Breit-Wigner total cross section for production of a $P_{cs}(4459)^0$ resonance with
spin $J=3/2$ in reaction (61) can be described on the basis of the spectral function (63),
provided that the branching fraction $Br[P_{cs}(4459)^0 \to {K^-}p]$ is known, as follows [44, 75--77]:
\begin{equation}
\sigma_{{K^-}p \to P_{cs}(4459)^0}(\sqrt{s^*},\Gamma_{cs})=4\left(\frac{\pi}{p^*_{K^-}}\right)^2
Br[P_{cs}(4459)^0 \to {K^-}p]S_{cs}(\sqrt{s^*},\Gamma_{cs})\Gamma_{cs}.
\end{equation}
Here, the center-of-mass 3-momentum of the colliding particles, $p^*_{K^-}$,
is defined above by Eq. (17). In line with [43], we employ in our cross-section calculations for the
branching ratio $Br[P_{cs}(4459)^0 \to {K^-}p]$ the value of 0.01\%, which was chosen in [43] to be
similar to that characterizing [78, 79] the $P_c^0 \to {\pi^-}p$ and $P_b^0 \to {\pi^-}p$ decays
\footnote{$^)$Thus, in the work [78] the obtained results show that the experimental data point for
the total cross section of the reaction ${\pi^-}p \to {J/\psi}n$ near the threshold is consistent with
the contribution from the hidden-charm pentaquark $P_{c}(4312)^0$ by assuming branching ratios
$Br[P_{c}(4312)^0 \to {J/\psi}n]$ $\approx$ 3\% and $Br[P_{c}(4312)^0 \to {\pi^-}p]$ $\approx$ 0.05\%.
Combined with the findings from the GlueX Collaboration relative to the branching fractions of the
decays of the $P_{c}(4312)^+$, $P_{c}(4440)^+$ and $P_{c}(4457)^+$ states to ${J/\psi}p$ [59], one can
assume that the value of about 0.05\% for the quantity $Br[P_{c}(4312)^0 \to {\pi^-}p]$ is quite reasonable.
Based on this assumption as well as on the similarity of the magnitudes of the total cross sections of
$K^-p$ and ${\pi^-}p$ scatterings [43], one can evaluate the branching ratio $Br[P_{cs}(4459)^0 \to {K^-}p]$
to be also around 0.05\%. The same value of 0.05\% was also used in Ref. [79] for the branching fraction
$Br[P_{b}^0 \to {\pi^-}p]$ in the study of the production of hidden-bottom pentaquark states $P_b^0$ in
${\pi^-}p$ interactions. Following Ref. [43], we chose in the present work more conservative value of 0.01\%
for the branching ratio $Br[P_{cs}(4459)^0 \to {K^-}p]$ to estimate the lower limit of the total cross section
for the $K^-p \to P_{cs}(4459)^0 \to {J/\psi}\Lambda$ reaction near threshold for given value of the
branching fraction $Br[P_{cs}(4459)^0 \to {J/\psi}\Lambda]$. Evidently, in case of use for the branching ratio
of the $P_{cs}(4459)^0$ decays to the $K^-p$ channel the aforesaid value of 0.05\% instead of that of 0.01\%,
adopted in the present work, the signal of $P_{cs}(4459)^0$ can be even more clearly distinguished from the
background.}$^)$
.
Within the representation of Eq. (64), the free total cross section
$\sigma_{{K^-}p \to P_{cs}(4459)^0\to {J/\psi}\Lambda}(\sqrt{s},\Gamma_{cs})$
for resonant $J/\psi$ production in the two-step process (61), (62) can be represented in the
following form [44]:
\begin{equation}
\sigma_{{K^-}p \to P_{cs}(4459)^0\to {J/\psi}\Lambda}(\sqrt{s},\Gamma_{cs})=
\sigma_{{K^-}p \to P_{cs}(4459)^0}(\sqrt{s},\Gamma_{cs})\theta[\sqrt{s}-(m_{J/\psi}+m_{\Lambda})]
Br[P_{cs}(4459)^0 \to {J/\psi}\Lambda].
\end{equation}
Here, $\theta(x)$ is the standard step function and the center-of-mass 3-momentum in the incoming
$K^-p$ channel, $p^*_{K^-}$, entering into Eq. (64), is defined above by the formula (17),
in which one has to make the replacement $E_t^2-p_t^2 \to m^2_p$ (and $s^* \to s$).
According to [44], majority of the $P_{cs}(4459)^0$ resonances, having vacuum total decay width
in their rest frames $\Gamma_{cs}=17.3$ MeV decay to ${J/\psi}\Lambda$ outside of the target
nuclei of interest. Taking into account both this fact and the results presented above by
Eqs. (3), (4), we get the following expression for the $J/\psi$ total production cross section
in $K^-A$ reactions from the production/decay sequence (61), (62):
\begin{equation}
\sigma_{{K^-}A\to {J/\psi}X}^{({\rm sec})}({\bf p}_{K^-})=\left(\frac{Z}{A}\right)
I_{V}[A,\sigma^{\rm in}_{{P_{cs}}N}]
\left<\sigma_{{K^-}p \to P_{cs}(4459)^0}({\bf p}_{K^-})\right>_A
Br[P_{cs}(4459)^0 \to {J/\psi}\Lambda],
\end{equation}
where
\begin{equation}
\left<\sigma_{{K^-}p \to P_{cs}(4459)^0}({\bf p}_{K^-})\right>_A
\end{equation}
$$
=\int\int
P_A({\bf p}_t,E)d{\bf p}_tdE
\sigma_{{K^-}p \to P_{cs}(4459)^0}(\sqrt{s^*},\Gamma_{cs})
\theta[\sqrt{s^*}-(m_{J/\psi}+m_{\Lambda})].
$$
The quantity $I_{V}[A,\sigma^{\rm in}_{P_{cs}N}]$ in Eq. (67) is defined above by Eq. (4), in which one
needs to make the substitution $\sigma \to \sigma^{\rm in}_{P_{cs}N}$
\footnote{$^)$Since the modulus $p_{K^-}$ of the incident beam momentum ${\bf p}_{K^-}$
of interest is substantially larger than the modulus $p_t$ of the struck target proton momentum ${\bf p}_t$
($p_{K^-} \sim 10$ GeV/c, $p_{t} \sim 250$ MeV/c), we assume, using the formula (4) for
the quantity $I_{V}[A,\sigma^{\rm in}_{P_{cs}N}]$ in Eq. (67), that the $P_{cs}(4459)^0$ momentum in laboratory
frame, equal to ${\bf p}_{K^-}$+ ${\bf p}_t$, is parallel to the initial antikaon momentum ${\bf p}_{K^-}$
and, hence, the $P_{cs}(4459)^0$ resonance moves in the nucleus essentially in the forward direction.}$^)$
.
Here, $\sigma^{\rm in}_{P_{cs}N}$ is the $P_{cs}(4459)^0$--nucleon inelastic total cross section,
averaged over proton and neutron targets:
\begin{equation}
\sigma^{\rm in}_{P_{cs}N}=\frac{Z}{A}\sigma^{\rm in}_{P_{cs}p}+\frac{N}{A}\sigma^{\rm in}_{P_{cs}n},
\end{equation}
where $\sigma^{\rm in}_{P_{cs}p}$ and $\sigma^{\rm in}_{P_{cs}n}$ are the inelastic total cross sections of the
free $P_{cs}(4459)^0{p}$ and $P_{cs}(4459)^0{n}$ interactions.
Our estimates, based on the $\Xi_c{\bar D}^*$ molecular scenario for the $P_{cs}(4459)^0$ strange pentaquark,
show that we can neglect quasielastic $P_{cs}(4459)^0{N}$ rescatterings in its way out of the target nucleus.
In this scenario, the cross sections $\sigma^{\rm in}_{P_{cs}p}$ and $\sigma^{\rm in}_{P_{cs}n}$
can be evaluated as:
\begin{equation}
\sigma^{\rm in}_{P_{cs}p} \approx \sigma^{\rm in}_{{\Xi_c^0}p}+\sigma^{\rm in}_{{\bar D}^{*0}p},\,\,\,
\sigma^{\rm in}_{P_{cs}n} \approx \sigma^{\rm in}_{{\Xi_c^0}n}+\sigma^{\rm in}_{{\bar D}^{*0}n}.
\end{equation}
Here, $\sigma^{\rm in}_{{\Xi_c^0}p({\Xi_c^0}n)}$ and $\sigma^{\rm in}_{{\bar D}^{*0}p({\bar D}^{*0}n)}$
are the inelastic total cross sections of the free
${\Xi_c^0}p({\Xi_c^0}n)$ and ${\bar D}^{*0}p({\bar D}^{*0}n)$ interactions, respectively.
In view of the similarity of interactions of ${\Xi_c^0}$ and ${\Xi^-}$, ${\bar D}^{*0}$ and ${\bar D}^{0}$
with nucleons due to their quark structures
(${\Xi_c^0}=|dsc>$, ${\Xi^-}=|dss>$, ${\bar D}^{*0}=|u{\bar c}>$, ${\bar D}^{0}=|u{\bar c}>$), we assume
that $\sigma^{\rm in}_{{\Xi_c^0}p({\Xi_c^0}n)} \approx \sigma^{\rm in}_{{\Xi^-}p({\Xi^-}n)}$,
$\sigma^{\rm in}_{{\bar D}^{*0}p({\bar D}^{*0}n)} \approx \sigma^{\rm in}_{{\bar D}^{0}p({\bar D}^{0}n)}$
(cf. [80]). The knowledge about the latter cross sections is scarce. Thus, for example, only very recently
the ALICE Collaboration made a first measurement of the two-particle momentum correlation function of the
$pD^-$ pairs in high-multiplicity $pp$ collisions at $\sqrt{s}=13$ TeV, which indicates the attractive
nature of the proton--$D^-$ interaction [81]. In our calculations we adopt for them
the following constants in the momentum regime above of 1 GeV/c of interest:
$\sigma^{\rm in}_{{\Xi^-}p}=12.7$ mb [82--84], $\sigma^{\rm in}_{{\bar D}^{0}p}=0$ [85],
$\sigma^{\rm in}_{{\Xi^-}n}=20.8$ mb [86] and
$\sigma^{\rm in}_{{\bar D}^{0}n}=\sigma_{{\bar D}^{0}n \to D^-p}=12$ mb [80, 85].
Using these values, we obtain that $\sigma^{\rm in}_{P_{cs}N} \approx 22.7$ mb for $^{12}$C$_6$ and
$\sigma^{\rm in}_{P_{cs}N} \approx 24.7$ mb for $^{184}$W$_{74}$. We will employ in our calculations
an arithmetic average of these results of 23.7 mb for the cross section $\sigma^{\rm in}_{P_{cs}N}$
for both considered target nuclei $^{12}$C$_6$ and $^{184}$W$_{74}$.

  Now consider the $J/\psi$ inclusive differential cross section arising from the production and
decay of intermediate resonance $P_{cs}(4459)^0$ in $K^-A$ collisions.
According to [57, 87], Eqs. (2), (66) and energy-momentum conservation in the production/decay chain
(61), (62), this cross section assumes the form:
\begin{equation}
\frac{d\sigma_{{K^-}A\to {J/\psi}X}^{({\rm sec})}({\bf p}_{K^-},{\bf p}_{J/\psi})}
{d{\bf p}_{J/\psi}}=\left(\frac{Z}{A}\right)
I_{V}[A,\sigma^{\rm in}_{{P_{cs}}N}]
\left<\frac{d\sigma_{{K^-}p \to P_{cs}(4459)^0 \to {J/\psi}\Lambda}({\bf p}_{K^-},{\bf p}_{J/\psi})}
{d{\bf p}_{J/\psi}}\right>_A,
\end{equation}
where
\begin{equation}
\left<\frac{d\sigma_{{K^-}p \to P_{cs}(4459)^0 \to {J/\psi}\Lambda}({\bf p}_{K^-},{\bf p}_{J/\psi})}
{d{\bf p}_{J/\psi}}\right>_A=
\int\int P_A({\bf p}_t,E)d{\bf p}_tdE
\left[\frac{d\sigma_{{K^-}p \to P_{cs}(4459)^0 \to {J/\psi}\Lambda}(\sqrt{s^*},{\bf p}_{J/\psi})}
{d{\bf p}_{J/\psi}}\right].
\end{equation}
and
\begin{equation}
\frac{d\sigma_{{K^-}p \to P_{cs}(4459)^0 \to {J/\psi}\Lambda}(\sqrt{s^*},{\bf p}_{J/\psi})}
{d{\bf p}_{J/\psi}}=\sigma_{{K^-}p \to P_{cs}(4459)^0}(\sqrt{s^*},\Gamma_{cs})
\theta[\sqrt{s^*}-(m_{J/\psi}+m_{\Lambda})]\times
\end{equation}
$$
\times
\frac{1}{\Gamma_{cs}(\sqrt{s^*},{\bf p}_{K^-})}\int d{\bf p}_{\Lambda}
\frac{d\Gamma_{P_{cs}(4459)^0 \to {J/\psi}\Lambda}(\sqrt{s^*},{\bf p}_{J/\psi},{\bf p}_{\Lambda})}
{d{\bf p}_{J/\psi}d{\bf p}_{\Lambda}},
$$
\begin{equation}
\frac{d\Gamma_{P_{cs}(4459)^0 \to {J/\psi}\Lambda}(\sqrt{s^*},{\bf p}_{J/\psi},{\bf p}_{\Lambda})}
{d{\bf p}_{J/\psi}d{\bf p}_{\Lambda}}=\frac{1}{2E_{cs}}\frac{1}{2J+1}|M_{P_{cs}(4459)^0 \to {J/\psi}\Lambda}|^2
(2\pi)^4\delta(E_{cs}-E_{J/\psi}-E_{\Lambda})\times
\end{equation}
$$
\times
\delta({\bf p}_{cs}-{\bf p}_{J/\psi}-{\bf p}_{\Lambda})\frac{1}{(2\pi)^3{2E_{J/\psi}}}
\frac{1}{(2\pi)^3{2E_{\Lambda}}},
$$
\begin{equation}
\Gamma_{cs}(\sqrt{s^*},{\bf p}_{K^-})=\Gamma_{cs}/\gamma_{cs},
\end{equation}
\begin{equation}
E_{cs}=E_{K^-}+E_t,\,\,\,\,\,{\bf p}_{cs}={\bf p}_{K^-}+{\bf p}_{t},\,\,\,\,\,\gamma_{cs}=E_{cs}/\sqrt{s^*}.
\end{equation}
Here, $E_{\Lambda}$ is the $\Lambda$ hyperon total energy ($E_{\Lambda}=\sqrt{m^2_{\Lambda}+{\bf p}^2_{\Lambda}}$)
and $|M_{P_{cs}(4459)^0 \to {J/\psi}\Lambda}|^2$ is summarized over spin states of primary and secondary particles
matrix element squared describing the decay (62). The decay amplitude $M_{P_{cs}(4459)^0 \to {J/\psi}\Lambda}$
is generally contributed by partial waves with different relative orbital angular momentum $L$ [75--77].
The kinematically allowed angular momenta $L$ for the ${J/\psi}\Lambda$ originating from the $P_{cs}(4459)^0$
decay can be estimated by using the constraint
 $\sqrt{L(L+1)} \le {p^{*}_{J/\psi}}R_{\rm str}$ [88, 89], where $p^{*}_{J/\psi}$ is
the $J/\psi$ c.m. momentum
\footnote{$^)$ We remind that in the case of off-shell target proton it is defined above by Eq. (18).}$^)$
and $R_{\rm str}/2$ is the strong interaction radius. Thus, for example, for $K^-$ beam momentum of 10 GeV/c,
belonging to the resonance region, and for the free target proton at rest we have $p^{*}_{J/\psi}=0.66$ GeV/c,
so that with a typical $R_{\rm str}$ of one fermi ${p^{*}_{J/\psi}}R_{\rm str}=3.3$, leading to $L=0,1,2$.
If $P_{cs}(4459)^0$ pentaquark has $J^P=(3/2)^-$ (preferred option, see above),
the allowed by the total angular momentum-parity conservation in its decay to the
${J/\psi}$$(1^-)$$\Lambda$$(1/2)^+$ values
of $L$ are 0 and 2. The decay process $P_{cs}(4459)^0 \to {J/\psi}\Lambda$ with $L=1$ is forbidden by the
parity conservation. Since we are mainly interested in the resonance $P_{cs}(4459)^0$ region, which
is not far from the ${J/\psi}\Lambda$ threshold,
we suppose here, analogously to the assumption that the hidden-charm pentaquarks $P_c(4312)^+(1/2)^-$,
$P_c(4440)^+(1/2)^-$ and $P_c(4457)^+(3/2)^-$ decays to ${J/\psi}p$
are dominated by the lowest partial waves with zero relative orbital angular momentum [77, 90--92],
that the $P_{cs}(4459)^0$ $(3/2)^-$ decays
to ${J/\psi}\Lambda$  are also dominated by the lowest partial wave with the relative
orbital angular momentum $L=0$ and the decay amplitude $M_{P_{cs}(4459)^0 \to {J/\psi}\Lambda}$
is determined by $s$-wave only. This means that the matrix element squared
$|M_{P_{cs}(4459)^0 \to {J/\psi}\Lambda}|^2$ results in an isotropic angular distribution of the
$P_{cs}(4459)^0 \to {J/\psi}\Lambda$ decay for the considered spin-parity assignment of the
the $P_{cs}(4459)^0$ state. By taking this fact into account as well as  integrating Eq. (73) over
the momenta ${\bf p}_{J/\psi}$ and ${\bf p}_{\Lambda}$ in the $P_{cs}(4459)^0$ rest frame, we can
get the following relation between $|M_{P_{cs}(4459)^0 \to {J/\psi}\Lambda}|^2$ and the partial
width $\Gamma_{P_{cs}(4459)^0 \to {J/\psi}\Lambda}$ of the $P_{cs}(4459)^0 \to {J/\psi}\Lambda$ decay:
\begin{equation}
\frac{1}{2J+1}\frac{|M_{P_{cs}(4459)^0 \to {J/\psi}\Lambda}|^2}{(2\pi)^2}=
\frac{2s^*}{\pi{p^*_{J/\psi}}}\Gamma_{P_{cs}(4459)^0 \to {J/\psi}\Lambda}.
\end{equation}
By using the relation (76) and accounting for the fact that $(\pi/\sqrt{s^*})p^*_{J/\psi}=I_2(s^*,m_{J/\psi},m_{\Lambda})$,
one finds that the expression (72) reduces to a simpler form (cf. Eq. (10)):
\begin{equation}
\frac{d\sigma_{{K^-}p \to P_{cs}(4459)^0 \to {J/\psi}\Lambda}(\sqrt{s^*},{\bf p}_{J/\psi})}
{d{\bf p}_{J/\psi}}=\sigma_{{K^-}p \to P_{cs}(4459)^0}(\sqrt{s^*},\Gamma_{cs})
\theta[\sqrt{s^*}-(m_{J/\psi}+m_{\Lambda})]\times
\end{equation}
$$
\times
\frac{1}{I_2(s^*,m_{J/\psi},m_{\Lambda})}Br[P_{cs}(4459)^0 \to {J/\psi}\Lambda]
\frac{1}{4E_{J/\psi}}\frac{1}{(\omega+E_t)}
\delta\left[\omega+E_t-\sqrt{m_{\Lambda}^2+({\bf Q}+{\bf p}_t)^2}\right],
$$
where the quantities $\omega$ and ${\bf Q}$ are defined by Eq. (13).
It will be adopted in our calculations of the $J/\psi$ momentum spectrum from the processes
(61), (62) in $K^-A$ collisions. Integrating the full inclusive differential cross section (70)
over the angular domain of ${\Delta}{\bf \Omega}_{J/\psi}$=$0^{\circ} \le \theta_{J/\psi} \le 20^{\circ}$,
and $0 \le \varphi_{J/\psi} \le 2{\pi}$ of our interest,
we can represent this spectrum, corresponding to that angular domain, in the following form (cf. Eq. (33)):
\begin{equation}
\frac{d\sigma_{{K^-}A\to {J/\psi}X}^{({\rm sec})}
(p_{K^-},p_{J/\psi})}{dp_{J/\psi}}=
\int\limits_{{\Delta}{\bf \Omega}_{J/\psi}}^{}d{\bf \Omega}_{J/\psi}
\frac{d\sigma_{{K^-}A\to {J/\psi}X}^{({\rm sec})}
({\bf p}_{K^-},{\bf p}_{J/\psi})}{d{\bf p}_{J/\psi}}p_{J/\psi}^2
\end{equation}
$$
=2{\pi}\left(\frac{Z}{A}\right)I_{V}[A,\sigma^{\rm in}_{{P_{cs}}N}]
\int\limits_{\cos20^{\circ}}^{1}d\cos{{\theta_{J/\psi}}}
\left<\frac{d\sigma_{{K^-}p \to P_{cs}(4459)^0 \to {J/\psi}{\Lambda}}(p_{K^-},
p_{J/\psi},\theta_{J/\psi})}{dp_{J/\psi}d{\bf \Omega}_{J/\psi}}\right>_A.
$$

Before the end of this subsection, we calculate the $J/\psi$ energy spectrum \\
$d\sigma_{{K^-}p \to P_{cs}(4459)^0 \to {J/\psi}{\Lambda}}[\sqrt{s},p_{J/\psi}]/ dE_{J/\psi}$
from the production/decay sequence (61), (62), proceeding on the free target proton at rest,
as a function of the $J/\psi$ total energy $E_{J/\psi}$ in addition to that from the non-resonant
${K^-}p \to {J/\psi}\Lambda$ reaction (see Eq. (59)). The energy-momentum conservation in the sequence
leads, as is easy to see, to the conclusion that the kinematical characteristics
(maximal value of the production angle in the laboratory system,
kinematically allowed momenta and energies in this system, their two different angular dependences)
of $J/\psi$ mesons produced in it and in this reaction are the same at given initial antikaon momentum.
The differential cross section
$d\sigma_{{K^-}p \to P_{cs}(4459)^0 \to {J/\psi}{\Lambda}}[\sqrt{s},{\bf p}_{J/\psi}]/ d{\bf p}_{J/\psi}$
can be obtained from more general one (77) in the limits: ${\bf p}_t \to 0$,
$E_t \to m_p$ and $s^* \to s$. The integration of this cross section over the angle $\theta_{J/\psi}$
between the momenta ${\bf p}_{K^-}$ and ${\bf p}_{J/\psi}$ when this angle varies within the
kinematically allowed angular range (34) with accounting for the relation
$d{\bf p}_{J/\psi}=p_{J/\psi}E_{J/\psi}dE_{J/\psi}d\Omega_{J/\psi}$ yields:
\begin{equation}
\frac{d\sigma_{{K^-}p \to P_{cs}(4459)^0 \to {J/\psi}{\Lambda}}[\sqrt{s},p_{J/\psi}]}{dE_{J/\psi}}=
2\pi
\int\limits_{\cos{\theta_{J/\psi}^{\rm max}}}^{1}d\cos{\theta_{J/\psi}}p_{J/\psi}E_{J/\psi}
\frac{d\sigma_{{K^-}p \to P_{cs}(4459)^0 \to {J/\psi}\Lambda}[\sqrt{s},{\bf p}_{J/\psi}]}{d{\bf p}_{J/\psi}}=
\end{equation}
$$
=
\sigma_{{K^-}p \to P_{cs}(4459)^0}(\sqrt{s},\Gamma_{cs})\theta[\sqrt{s}-(m_{J/\psi}+m_{\Lambda})]\times
$$
$$
\times
\left(\frac{\sqrt{s}}{2p_{K^-}p^{*}_{J/\psi}}\right)
Br[P_{cs}(4459)^0 \to {J/\psi}\Lambda]~{\rm for}
~E^{(2)}_{J/\psi}(0^{\circ}) \le E_{J/\psi} \le E^{(1)}_{J/\psi}(0^{\circ}).
$$
Eq. (79) evidently shows that the $J/\psi$ energy spectrum, which arises from the free space
production/decay chain (61), (62), exhibits a characteristic stepwise (flat) behavior within the
allowed energy range (58), where its dependence on $E_{J/\psi}$ is completely absent.
\begin{figure}[htb]
\begin{center}
\includegraphics[width=16.0cm]{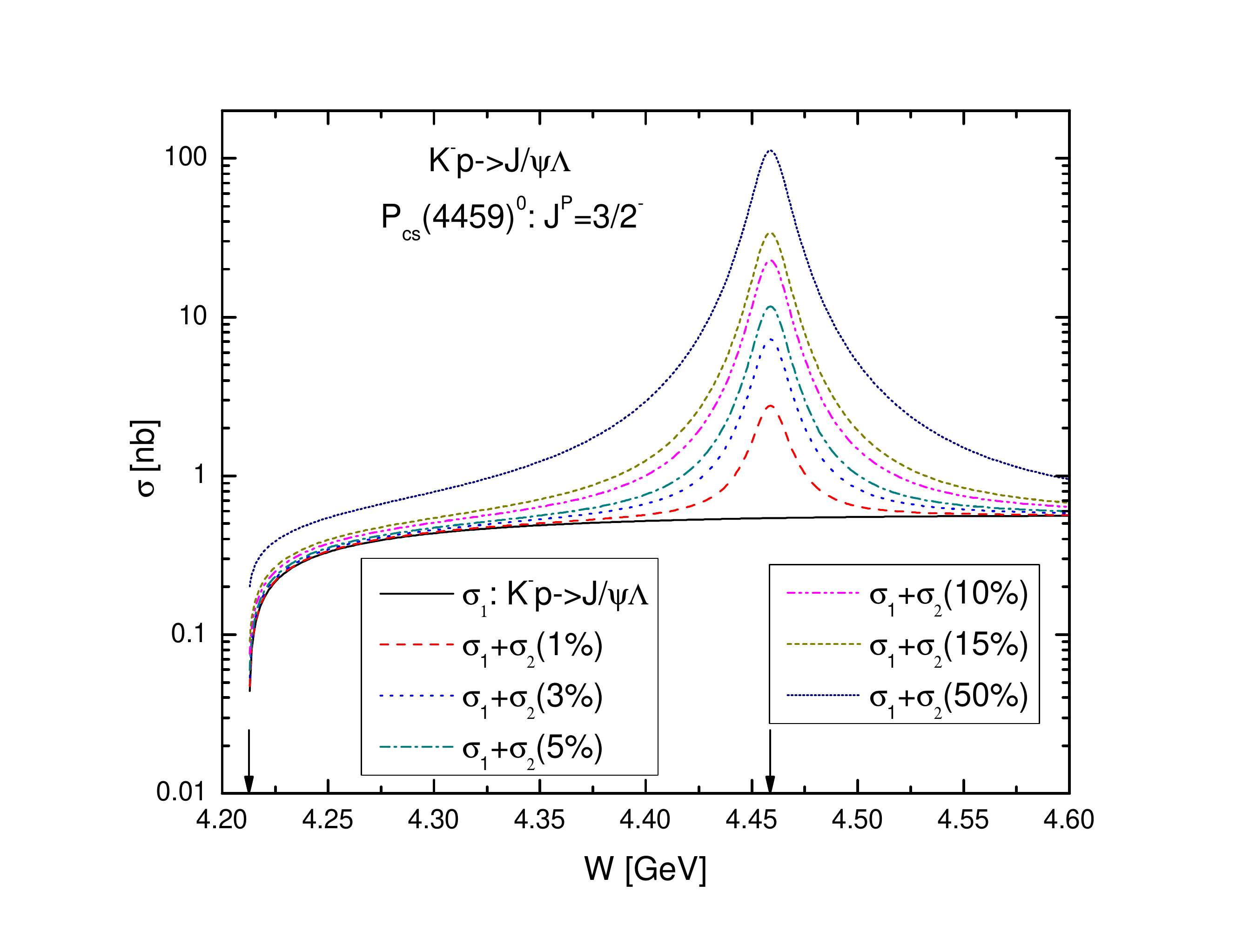}
\vspace*{-2mm} \caption{(Color online) The non-resonant total cross section for the reaction
${K^-}p \to {J/\psi}\Lambda$ (solid curve), calculated on the basis of Eqs. (29)--(32).
Incoherent sum of it and the total cross section for the
resonant $J/\psi$ production in the process ${K^-}p \to P_{cs}(4459)^0 \to {J/\psi}\Lambda$,
calculated in line with Eq. (65) assuming that the resonance $P_{cs}(4459)^0$
has the spin-parity quantum numbers $J^P=(3/2)^-$ and decays to $K^-p$ and ${J/\psi}\Lambda$
with branching fractions 0.01\% and 1, 3, 5, 10, 15, 50\%
(respectively, dashed, dotted, dashed-dotted, dashed-dotted-dotted, short-dashed and short-dotted curves),
as functions of the center-of-mass energy $W=\sqrt{s}$ of the antikaon--proton collisions.
The left and right arrows indicate, correspondingly, the center-of-mass threshold energy of 4.2126 GeV
for direct $J/\psi$ production on a free target proton being at rest and
the resonant energy of 4.4588 GeV.}
\label{void}
\end{center}
\end{figure}
\begin{figure}[!h]
\begin{center}
\includegraphics[width=16.0cm]{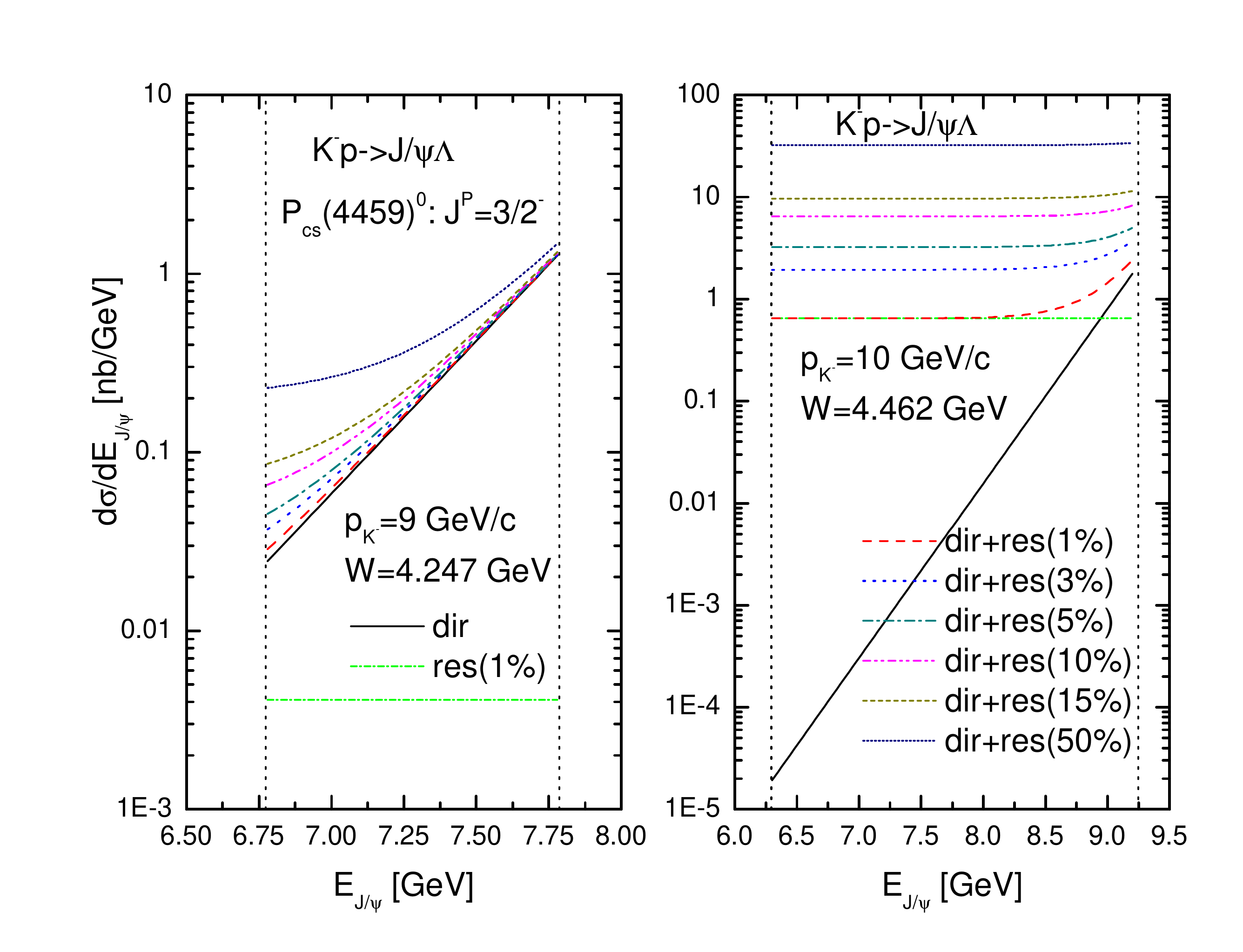}
\vspace*{-2mm} \caption{(Color online) The non-resonant $J/\psi$ energy spectrum in the reaction
${K^-}p \to {J/\psi}\Lambda$, calculated on the basis of Eq. (59) at incident
antikaon momenta of 9 (left panel) and 10 GeV/c (right panel)
in the laboratory system (respectively, solid curves in the left and right panels).
The resonant $J/\psi$ energy spectrum in the process ${K^-}p \to P_{cs}(4459)^0 \to {J/\psi}\Lambda$,
calculated in line with Eq. (79) at the same incident antikaon momenta of 9 and 10 GeV/c as above,
assuming that the resonance $P_{cs}(4459)^0$ with the spin-parity quantum numbers $J^P=(3/2)^-$ decays to ${J/\psi}\Lambda$ with the lower allowed relative orbital angular momentum $L=0$ with branching fraction 1\% (respectively, short-dashed-dotted curves in the left and right panels).
Incoherent sum of the non-resonant $J/\psi$ energy spectrum and resonant one, calculated assuming that the resonance $P_{cs}(4459)^0$ with the spin-parity combination $J^P=(3/2)^-$ decays to ${J/\psi}\Lambda$ with the lower allowed relative orbital angular momentum $L=0$ with branching fractions 1, 3, 5, 10, 15 and 50\% (respectively, dashed, dotted, dashed-dotted, dashed-dotted-dotted, short-dashed and short-dotted curves in the left and right panels), all as functions of the total $J/\psi$ energy $E_{J/\psi}$ in the laboratory frame.
The two vertical dotted lines in the left and right panels mark the range of kinematically allowed energies of $J/\psi$ mesons in this frame for the considered direct and resonant $J/\psi$ production on a free target proton being at rest at
given initial antikaon momenta of 9 and 10 GeV/c, respectively.}
\label{void}
\end{center}
\end{figure}
\begin{figure}[!h]
\begin{center}
\includegraphics[width=16.0cm]{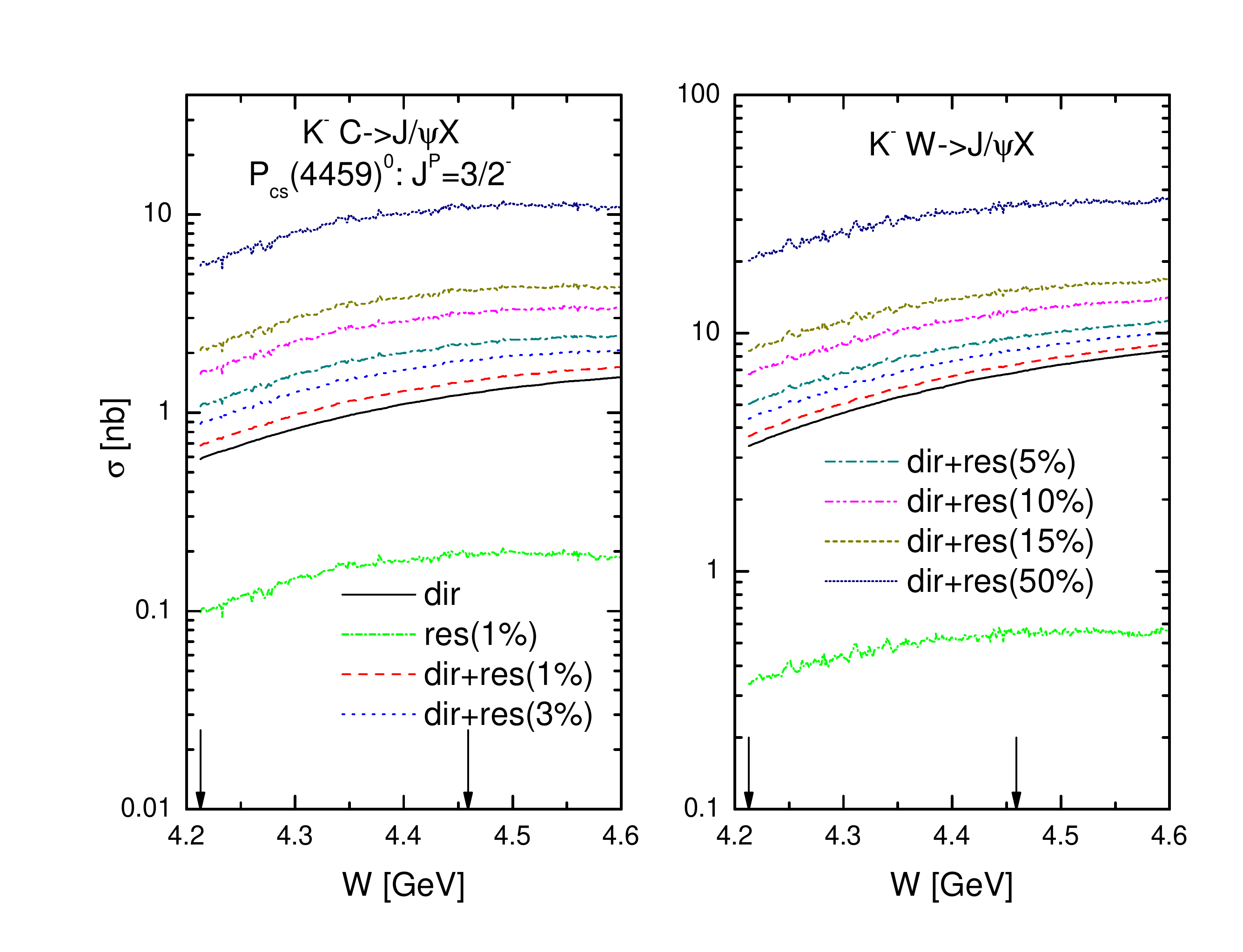}
\vspace*{-2mm} \caption{(Color online) The non-resonant $J/\psi$ production total cross sections in the reactions
${K^-}{\rm ^{12}C} \to {J/\psi}X$ (left panel) and ${K^-}{\rm ^{184}W} \to {J/\psi}X$ (right panel)
(solid curves in the left and right panels), calculated on the basis of Eq. (3).
The resonant $J/\psi$ production total cross sections in the process ${K^-}p \to P_{cs}(4459)^0 \to {J/\psi}\Lambda$,
proceeding on the intranuclear protons of carbon and tungsten target nuclei.
They were obtained in line with Eq. (66), assuming that the resonance $P_{cs}(4459)^0$ with the spin-parity quantum numbers $J^P=(3/2)^-$ decays to ${J/\psi}\Lambda$ with the lower allowed relative orbital angular momentum $L=0$ with branching fraction 1\% (respectively, short-dashed-dotted curves in the left and right panels). Incoherent sum of the non-resonant $J/\psi$ production total cross sections and resonant ones, calculated assuming that the resonance $P_{cs}(4459)^0$ with the spin-parity configuration $J^P=(3/2)^-$ decays to ${J/\psi}\Lambda$
with the lower allowed relative orbital angular momentum $L=0$ with branching fractions 1, 3, 5, 10, 15 and 50\% (respectively, dashed, dotted, dashed-dotted, dashed-dotted-dotted, short-dashed and short-dotted curves in the left and right panels), all as functions of the center-of-mass energy $W=\sqrt{s}$ of the free antikaon--proton collisions.
The two arrows in both panels indicate the center-of-mass threshold energy of 4.2126 GeV
for direct $J/\psi$ production on a free target proton being at rest and
the resonant energy of 4.4588 GeV.}
\label{void}
\end{center}
\end{figure}
\begin{figure}[!h]
\begin{center}
\includegraphics[width=16.0cm]{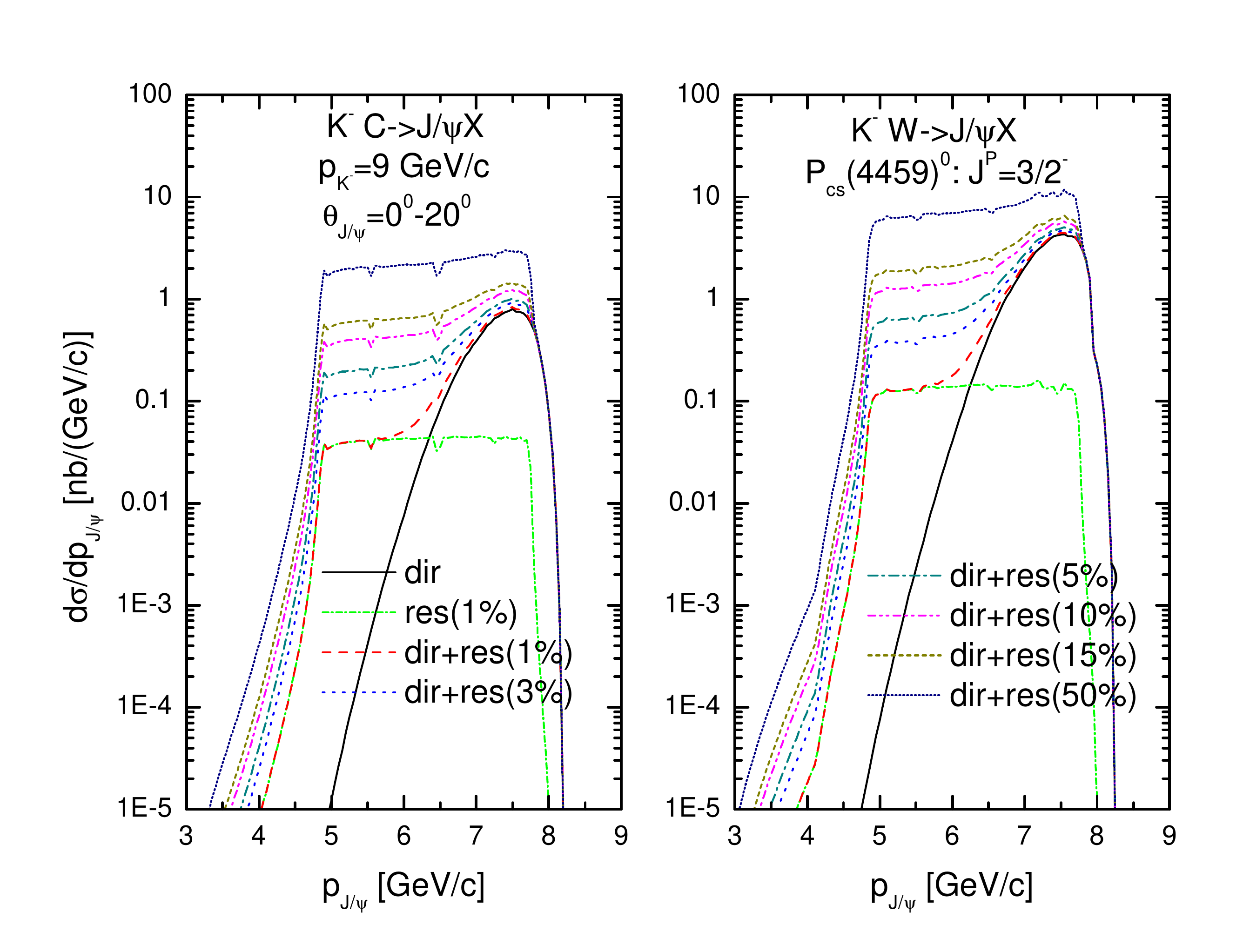}
\vspace*{-2mm} \caption{(Color online) The non-resonant momentum differential cross section for the
production of $J/\psi$ mesons in the reaction ${K^-}{\rm ^{12}C} \to {J/\psi}X$ (left panel) and
${K^-}{\rm ^{184}W} \to {J/\psi}X$ (right panel) in the laboratory polar angular range of 0$^{\circ}$--20$^{\circ}$, calculated on the basis of Eq. (33) at incident antikaon momentum of 9 GeV/c in the laboratory system (solid curves in the left and right panels).
The resonant momentum differential cross section for the production of $J/\psi$ mesons
in the two-step process ${K^-}p \to P_{cs}(4459)^0 \to {J/\psi}\Lambda$,
proceeding on the intranuclear protons of carbon and
tungsten target nuclei, in the laboratory polar angular range of 0$^{\circ}$--20$^{\circ}$.
It was calculated in line with Eq. (78) at the same incident antikaon momentum of 9 GeV/c
as above assuming that the resonance $P_{cs}(4459)^0$ with the spin-parity quantum numbers $J^P=(3/2)^-$
decays to ${J/\psi}\Lambda$ with the lower allowed relative orbital angular momentum $L=0$ with branching fraction
1\% (short-dashed-dotted curves in the left and right panels). Incoherent sum of the non-resonant
$J/\psi$ momentum differential cross section and resonant one, calculated assuming that the resonance $P_{cs}(4459)^0$ with the spin-parity quantum numbers $J^P=(3/2)^-$ decays to ${J/\psi}\Lambda$
with the lower allowed relative orbital angular momentum $L=0$ with branching fractions 1, 3, 5, 10, 15 and 50\% (respectively, dashed, dotted, dashed-dotted, dashed-dotted-dotted, short-dashed and short-dotted curves in the left and right panels), all as functions of the $J/\psi$ momentum $p_{J/\psi}$ in the laboratory frame.}
\label{void}
\end{center}
\end{figure}
\begin{figure}[htb]
\begin{center}
\includegraphics[width=16.0cm]{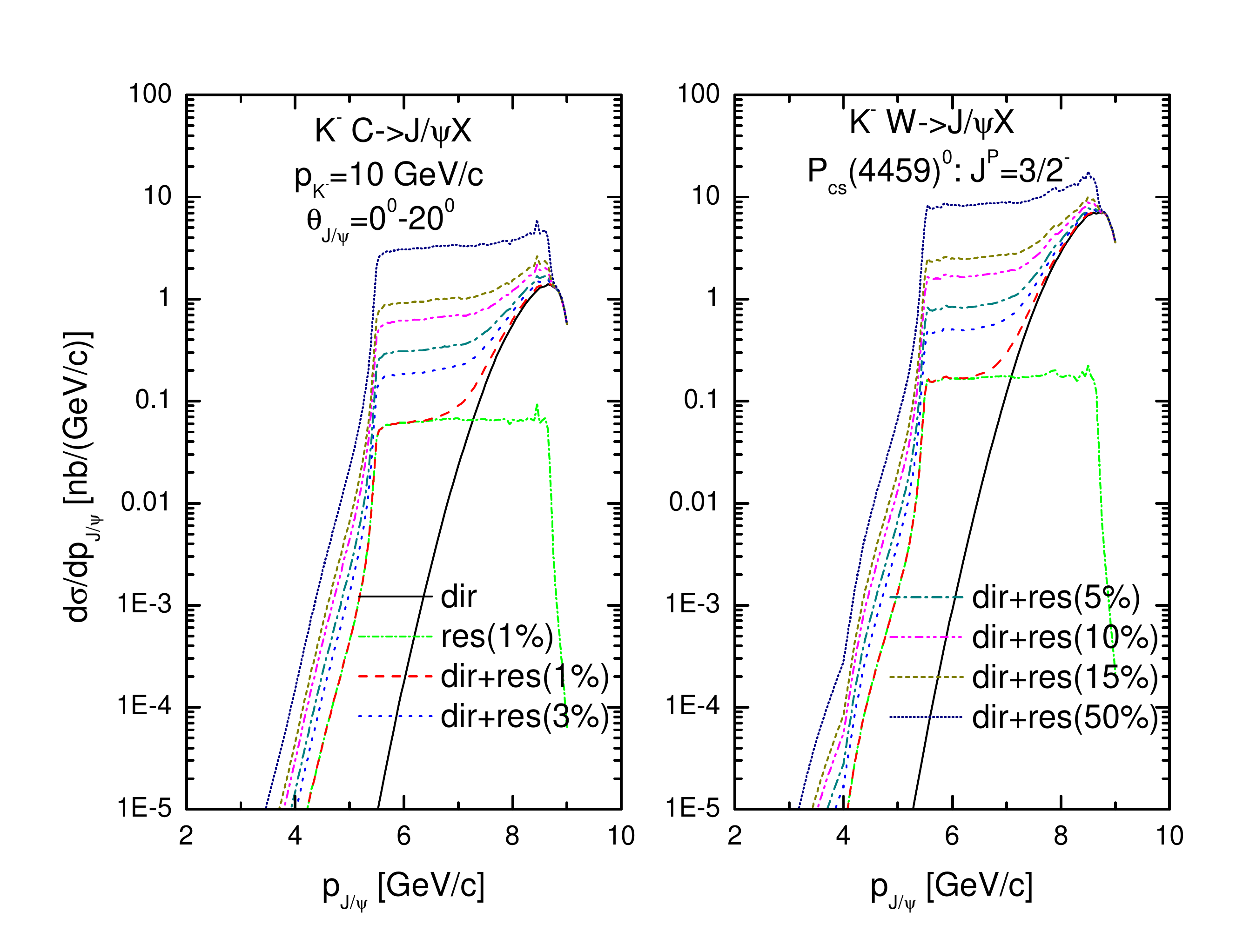}
\vspace*{-2mm} \caption{(Color online) The same as in Fig. 8, but for the initial antikaon momentum of 10 GeV/c.}
\label{void}
\end{center}
\end{figure}

\section*{3. Results and discussion}

  The vacuum elementary non-resonant $J/\psi$ production cross section in the reaction
${K^-}p \to {J/\psi}\Lambda$, determined on the basis of Eqs. (29)--(32),
and the combined (non-resonant plus resonant (65)) $J/\psi$ production total cross section are depicted in Fig. 5
for the considered spin-parity assignment of the hidden-charm strange resonance $P_{cs}(4459)^0$.
From this figure one can see that the $P_{cs}(4459)^0$ state appears as distinct peak at $W=$ 4.4588 GeV
in the combined cross section, if $Br[P_{cs}(4459)^0 \to {J/\psi}\Lambda]=3$, 5, 10, 15 and 50\%.
In these cases, at c.m.s. antikaon energies around the peak energy the resonant
contributions are much larger than the non-resonant ones of about 1 nb. Therefore, the background reaction will
not influence the direct observation of the hidden-charm strange pentaquark $P_{cs}(4459)^0$
production at these energies and in these cases. The peak values of the combined cross section reach
a well measurable values $\sim$ 10 and 100 nb, if $Br[P_{cs}(4459)^0 \to {J/\psi}\Lambda]=3$ and 50\%,
correspondingly. If $Br[P_{cs}(4459)^0 \to {J/\psi}\Lambda]=$~1\%, then
the resonant and non-resonant $J/\psi$ yields are comparable in the resonance region.
It should be pointed out that the peak strengths of the combined cross section
of the reaction ${K^-}p \to {J/\psi}\Lambda$ for branching fractions
$Br[P_{cs}(4459)^0 \to {J/\psi}\Lambda]=1$, 10 and 50\% are similar to those determined in the recent work [43].
In view of the above, it is natural to expect that the $P_{cs}(4459)^0$ signal could be well distinguished from the background reaction, if branching ratio $Br[P_{cs}(4459)^0 \to {J/\psi}\Lambda]$ $\sim$ 3\% and more.
To see experimentally such peak structure in the combined total cross section of the reaction
${K^-}p \to {J/\psi}\Lambda$
\footnote{$^)$The $J/\psi$ mesons could be identified via the decays $J/\psi \to {\mu^+}{\mu^-}$
or $J/\psi \to {e^+}{e^-}$ with a branching ratios of about 6\% [93].}$^)$,
it is enough to have the antikaon momentum resolution (and the momentum binning) of the order of 10 MeV/c.
Thus, the c.m. energy range
$M_{cs}-{\Gamma_{cs}}/2 < \sqrt{s} < M_{cs}+{\Gamma_{cs}}/2$ corresponds to the laboratory
antikaon momentum region of 9.942 GeV/c $< p_{K^-} <$ 10.024 GeV/c,
i.e. ${\Delta}p_{K^-}=$ 82 MeV/c for the $P_{cs}(4459)^0$.
This means that to resolve the peak in Fig. 5 the $K^-$ beam momentum resolution
(and the momentum bin size) of the order of 10 MeV/c are required. This requirement is expected to be
satisfied at the K10 beam line at the J-PARC Hadron Experimental Facility, where the antikaon incident
momentum resolution is assumed to be $\Delta{p}_{beam}/p_{beam}=0.1$\% or 10 MeV/c for initial $K^-$ momentum
of 10 GeV/c [41, 42]. Therefore, the high-intensity, high-momentum and high-resolution $K^-$ beam, which will be
available at this beam line, will allow to study the pentaquark state $P_{cs}(4459)^0$ in the  reaction
$K^-p \to {J/\psi}\Lambda$.

  The model described above makes it possible to calculate the non-resonant $J/\psi$ energy spectrum
from the considered ${K^-}p \to {J/\psi}\Lambda$ elementary reaction, the resonant one
from the production/decay sequence (61), (62), proceeding on the free target proton being at rest.
They were calculated according to Eqs. (59), (79), respectively, for incident antikaon momenta of 9 and 10 GeV/c.
The resonant $J/\psi$ energy spectrum was obtained for the considered spin-parity assignment
of the $P_{cs}(4459)^0$ resonance for branching fraction $Br[P_{cs}(4459)^0 \to {J/\psi}\Lambda]=$~1\%.
These dependences, together with the incoherent sum of the non-resonant $J/\psi$ energy spectrum and resonant one,
calculated for six employed scenarios for the branching ratio of the decay $P_{cs}(4459)^0 \to {J/\psi}\Lambda$,
as functions of the $J/\psi$ total energy $E_{J/\psi}$ are shown, respectively, in left and right panels of Fig. 6.
One can see from this figure that while the resonant $J/\psi$ production cross section shows a flat behavior
at all allowed total energies $E_{J/\psi}$, the non-resonant cross section falls off quickly as $E_{J/\psi}$
decreases. At near-threshold incident $K^-$ beam momentum of 9 GeV/c its strength is substantially larger than
that of the resonant $J/\psi$ production cross section, calculated for the value of the branching ratio
$Br[P_{cs}(4459)^0 \to {J/\psi}\Lambda]=$~1\% for all allowed $J/\psi$ total energies. Thus, for example, in this case
for the $J/\psi$ mesons with total energy of 7.0 GeV their non-resonant production cross section is enhanced compared
to the resonant one by a factor of about 15. As a result, the $J/\psi$ meson combined differential energy distribution
reveals some sensitivity to the employed variations in the branching ratio of the
$P_{cs}(4459)^0 \to {J/\psi}\Lambda$ decays only at "low" (at 6.8--7.0 GeV) $J/\psi$ total energies.
Here, the $J/\psi$ combined yield is enhanced for the values of this ratio of 3, 5, 10, 15 and 50\% by sizeable
factors of about 1.3, 1.6, 2.2, 2.8 and 7.0, respectively, as compared to that from the directly produced $J/\psi$ mesons. If $Br[P_{cs}(4459)^0 \to {J/\psi}\Lambda]=$~1\%, then the
combined $J/\psi$ yield is practically indistinguishable from the non-resonant background for all allowed $J/\psi$
total energies. This implies that in the case of the production and decay of the intermediate $P_{cs}(4459)^0$ $(3/2)^-$ resonance the differences between the combined results obtained by using a conservative values of the branching fraction of the decays $P_{cs}(4459)^0 \to {J/\psi}\Lambda$ of 1 and 3\%, 3 and 5\%, 5 and 10\%, 10 and 15\% are also sizeable and
experimentally measurable. They are $\sim$ 20--30\%. On the other hand, in the $p_{K^-}=10$ GeV/c case, the combined yield is entirely caused by the presence of the $P_{cs}(4459)^0$ resonance in $J/\psi$ production and shows, contrary to the previous case, practically flat characteristic behavior at $J/\psi$ total energies below 9.0 GeV. Its strength is strictly proportional to the branching fraction $Br[P_{cs}(4459)^0 \to {J/\psi}\Lambda]$. As a result, the differences between all calculations corresponding to the adopted options for this fraction, are quite visible and experimentally distinguishable.
Therefore, the $J/\psi$ meson energy differential cross sections measurements in the $K^-p \to {J/\psi}\Lambda$ reaction not only in the resonance, but also in the close to threshold incident $K^-$ meson momentum regions in future high-precision experiments at the J-PARC should provide both further evidence for the existence of the hidden-charm strange pentaquark $P_{cs}(4459)^0$ state and also clarify its decay rate. Since the $J/\psi$ production differential cross sections at beam momentum of 10 GeV/c have a well measurable absolute values $\sim$ 1--30 nb/GeV and they
are larger than those at antikaon momentum of 9 GeV/c and for "low" $J/\psi$ energies by about of one to two orders of magnitude (compare left and right panels of Fig. 6), the measurements in the resonance region are evidently favored compared to those in the close to threshold domain.

    Excitation functions for the non-resonant production of $J/\psi$ mesons, for
their resonant production via $P_{cs}(4459)^0$ resonance formation and its decay to ${J/\psi}\Lambda$
with branching ratio $Br[P_{cs}(4459)^0 \to {J/\psi}\Lambda]=$~1\% as well as for the combined
(non-resonant+resonant) $J/\psi$ production in ${K^-}$$^{12}$C and $K^-$$^{184}$W collisions
are given, respectively, in the left and right panels of Fig.~7. The latter ones are calculated using
Eqs.~(3), (66) and assuming that branching ratio
$Br[P_{cs}(4459)^0 \to {J/\psi}\Lambda]=1$, 3, 5, 10, 15 and 50\%.
One can see that the resonant $J/\psi$ production, contrary to the case of the target proton being at rest
(cf. Fig. 5), is smeared out by the Fermi motion of intranuclear protons. Its strength, calculated for the
branching fraction $Br[P_{cs}(4459)^0 \to {J/\psi}\Lambda]=$~1\%, is reduced by a factor of about ten as
compared to that of the non-resonant production for both considered target nuclei.
Nevertheless, there are a sizeable distinctions between the results for the combined total $J/\psi$
production cross section corresponding to the considered conservative
choices 1 and 3\%, 3 and 5\%, 5 and 10\%, 10 and 15\% for the branching ratio $Br[P_{cs}(4459)^0 \to {J/\psi}\Lambda]$
at the c.m.s. beam energies of interest. They are $\sim$ 25--30\% for the carbon nucleus. In the case of a tungsten
target nucleus these distinctions are somewhat lower, they are $\sim$ 15--20\%. The additional enhancement
in the behavior of the total $J/\psi$ production cross section on both target nuclei,
produced by the $P_{cs}(4459)^0 \to {J/\psi}\Lambda$ decays with the branching ratio of 1\%, is small. It is about
10\%. This means that, if the branching fraction $Br[P_{cs}(4459)^0 \to {J/\psi}\Lambda]$ $\sim$ 3\% and more, it can
be accurately studied experimentally in the dedicated experiment at the J-PARC through the charmonium excitation
function measurements not only on the proton target (cf. Fig. 5), but also on the nuclear targets near the threshold
as well as the comparison of their results with the calculations based on the present model with known total cross
section of the direct process (1). Since the $J/\psi$ meson production cross sections on $^{184}$W nucleus are larger
than those on the $^{12}$C by a factor of about five, the collected statistics in the measurements on heavy target
nuclei are expected to be substantially higher than what could be achieved in measurements on the light nuclear
targets. This should enable an accurate determination of the above fraction in the measurements both on
light and heavy nuclear targets. Evidently, the advantage of these measurements compared to those on the proton
target is that they could be performed with moderate $K^-$ beam momentum resolution and, hence, there is no need
to undertake in them a detailed scan of the nuclear $J/\psi$ total production cross section in the near-threshold region
to provide further evidence for the existence of the hidden-charm pentaquark $P_{cs}(4459)^0$ with
strangeness and to obtain valuable information on its decay rate to the ${J/\psi}\Lambda$ mode.

     The momentum dependences of the absolute non-resonant, resonant and combined $J/\psi$ meson differential
cross sections, respectively, from the direct (1), two-step (61), (62) and direct plus two-step $J/\psi$
production processes in $K^-$$^{12}$C and $K^-$$^{184}$W interactions, calculated on the basis of
Eqs. (33), (78) for laboratory angles of 0$^{\circ}$--20$^{\circ}$ and for initial antikaon momenta
of 9 and 10 GeV/c, are depicted, correspondingly, in Figs. 8 and 9. The resonant momentum differential cross
section for the production of $J/\psi$ mesons in the two-step process
${K^-}p \to P_{cs}(4459)^0 \to {J/\psi}\Lambda$,
proceeding on the intranuclear protons of carbon and tungsten target nuclei was obtained
for six adopted values of the branching fraction $Br[P_{cs}(4459)^0 \to {J/\psi}\Lambda]$.
It is seen from these figures that the contribution to the $J/\psi$ production on both these nuclei, which comes
from the intermediate $P_{cs}(4459)^0$ state, is much larger than that from the background process (1)
in the "low"-momentum regions of 4--6 GeV/c and 4--7 GeV/c for considered $K^-$ beam momenta of 9 and 10 GeV/c,
respectively. In them, the combined yield is entirely governed by the presence of the $P_{cs}(4459)^0$ state
in $J/\psi$ production and shows practically flat behavior. Its strength is strictly proportional to the branching ratio $Br[P_{cs}(4459)^0 \to {J/\psi}\Lambda]$ used in the calculations with a value increasing by a factor of about 3
for both $K^-$ beam momenta considered, when going from carbon target to tungsten one.
As a result, the differences between all calculations corresponding to the employed options for this ratio, are well separated and experimentally distinguishable. Since the ratios between the differential cross sections for the
production of $J/\psi$ mesons by 10 GeV/c $K^-$ mesons on $^{12}$C and $^{184}$W nuclei, and
the cross sections for their production on these target nuclei by 9 GeV/c antikaons
in the above "low"-momentum regions (plateau regions) are only about 1.5,
this should enable an accurate determination of this ratio -- at least to distinguish between its conservative
options of 1, 3, 5, 10 and 15\% -- also in the $J/\psi$ meson momentum differential
cross section measurements on light and especially on heavy nuclear targets not only in the resonance region
(at momenta of incoming $K^-$ mesons around 10 GeV/c), but also at their momenta belonging to the threshold region (at momenta $\sim$ 9 GeV/c). Such measurements could be performed in the future at the J-PARC Hadron Experimental Facility
using the high-intensity separated secondary $K^-$ beams.

  Taking into account the above considerations, we come to the conclusion that the near-threshold $J/\psi$
excitation function, energy and momentum distribution measurements in antikaon-induced reactions both on protons
and on nuclear targets will provide further evidence for the existence of the pentaquark
$P_{cs}(4459)^0$ resonance and will shed light on its decay rate to the channel ${J/\psi}\Lambda$.

\section*{4. Conclusions}

 In this paper we have calculated the absolute excitation functions for the non-resonant, resonant and for the
combined (non-resonant plus resonant) production of $J/\psi$ mesons off protons at incident antikaon c.m.s.
excess energies above the lowest ${J/\psi}\Lambda$ production threshold
below 0.4 GeV by considering direct non-resonant $K^-p \to {J/\psi}\Lambda$ and two-step resonant
${K^-}p \to P_{cs}(4459)^0 \to {J/\psi}\Lambda$ $J/\psi$ production channels as well as assuming the spin-parity assignment of the hidden-charm resonance $P_{cs}(4459)^0$ with strangeness as $J^P=(3/2)^-$
within six different scenarios for the branching ratio $Br[P_{cs}(4459)^0 \to {J/\psi}\Lambda]$
of the decay $P_{cs}(4459)^0 \to {J/\psi}\Lambda$,
namely: $Br[P_{cs}(4459)^0 \to {J/\psi}\Lambda]=$~1, 3, 5, 10, 15 and 50\%.
Also, an analogous functions for the production of $J/\psi$ mesons on $^{12}$C and $^{184}$W
target nuclei at the same near-threshold center-of-mass beam energies as in the case of proton target
have been calculated by considering aforementioned incoherent direct and two-step
$J/\psi$ production processes within a nuclear spectral function approach.
Furthermore, the non-resonant $J/\psi$ energy spectrum from the ${K^-}p \to {J/\psi}\Lambda$
elementary reaction, the resonant one from the production/decay sequence
${K^-}p \to P_{cs}(4459)^0 \to {J/\psi}\Lambda$, proceeding on the free target proton being at rest,
and the incoherent sum of the non-resonant $J/\psi$ energy spectrum and resonant one
were calculated for the considered spin-parity assignment of the $P_{cs}(4459)^0$ resonance and for
the adopted branching fractions of its decay to the ${J/\psi}\Lambda$ final state at incident antikaon
momenta of 9 and 10 GeV/c. In addition to this, the momentum dependences of the absolute non-resonant,
resonant and combined $J/\psi$ meson differential cross sections from the considered direct,
two-step and direct plus two-step $J/\psi$ production elementary processes in $K^-$$^{12}$C and $K^-$$^{184}$W
interactions were obtained for laboratory angles of 0$^{\circ}$--20$^{\circ}$ and for initial $K^-$
momenta of 9 and 10 GeV/c as well. The combined momentum differential cross
sections for the production of $J/\psi$ mesons in the direct and two-step processes were determined
for six adopted values of the branching fraction $Br[P_{cs}(4459)^0 \to {J/\psi}\Lambda]$.
It was shown that the $P_{cs}(4459)^0$ state appears as clear narrow independent peak at c.m.s. energy
$W=$~4.4588 GeV in the combined cross section on proton target,
if $Br[P_{cs}(4459)^0 \to {J/\psi}\Lambda]=3$, 5, 10, 15 and 50\%.
The peak values of this cross section reach ten and hundred of nanobarns,
if $Br[P_{cs}(4459)^0 \to {J/\psi}\Lambda]=3$ and 50\%, respectively. Therefore, a detailed scan of the
$J/\psi$ total production cross section on a proton target in antikaon-induced reactions
in the near-threshold energy region in future high-precision experiments, for example, at the J-PARC
should give a further evidence for the existence of the hidden-charm strange pentaquark state
$P_{cs}(4459)^0$ and clarify its decay rate to the ${J/\psi}\Lambda$ channel.
It was also demonstrated that the presence of the $P_{cs}(4459)^0$ pentaquark resonance in
$J/\psi$ production on nuclei in $K^-$$^{12}$C and $K^-$$^{184}$W collisions leads to
a sizeable and experimentally measurable differences ($\sim$ 25--30\% for $^{12}$C and $\sim$ 15--20\% for $^{184}$W)
between the results for the combined total $J/\psi$ production cross section on these nuclei corresponding to the considered conservative choices 1 and 3\%, 3 and 5\%, 5 and 10\%, 10 and 15\% for the branching ratio
$Br[P_{cs}(4459)^0 \to {J/\psi}\Lambda]$ at all c.m.s. beam energies of interest.
The additional enhancement in the behavior of the total $J/\psi$ production cross section on both target nuclei,
produced by the $P_{cs}(4459)^0 \to {J/\psi}\Lambda$ decays with the branching ratio of 1\%, is small. It is about
10\%. This offers an indirect possibility of studying of the branching fraction $Br[P_{cs}(4459)^0 \to {J/\psi}\Lambda]$ experimentally in the future dedicated experiment at the J-PARC via the near-threshold charmonium excitation function measurements not only on the proton target, but also on the nuclear targets, if it $\sim$ 3\% and more.
It was further shown that also the near-threshold $J/\psi$ energy and
momentum distribution measurements in antikaon-induced reactions, respectively, on protons
and on nuclear targets will provide further convincing evidence for the existence of the pentaquark
$P_{cs}(4459)^0$ resonance and will shed light on its decay rate to the ${J/\psi}\Lambda$ final state
-- at least will help to distinguish between its conservative (and realistic) options of 1, 3, 5, 10 and 15\%.
Such measurements could be performed in the future at the J-PARC Hadron Experimental Facility as well.


\begin{thebibliography}{99}
\bibitem{1} J. J. Wu, R. Molina, E. Oset and B. S. Zou, Phys. Rev. Lett. {\bf 105}, 232001 (2010);\\                                   arXiv:1007.0573 [nucl-th].
\bibitem{2} J. J. Wu, R. Molina, E. Oset and B. S. Zou, Phys. Rev. C {\bf 84}, 015202 (2011);\\                                   arXiv:1011.2399 [nucl-th].
\bibitem{3} W. L. Wang, F. Huang, Z. Y. Zhang and B. S. Zou, Phys. Rev. C {\bf 84}, 015203 (2011);\\
              arXiv:1101.0453 [nucl-th].
\bibitem{4} J. J. Wu, T.-S. H. Lee and B. S. Zou, Phys. Rev. C {\bf 85}, 044002 (2012);\\
               arXiv:1202.1036 [nucl-th].
\bibitem{5} Z. C. Yang, Z. F. Sun, J. He, X. Liu and S.-L. Zhu, Chin. Phys. C {\bf 36}, 6 (2012);\\
               arXiv:1105.2901 [hep-ph].
\bibitem{6} C. Garcia-Recio, J. Nieves, O. Romanets, L. L. Salcedo and
              L. Tolos, Phys. Rev. D {\bf 87}, 074034 (2013);\\
               arXiv:1302.6938 [hep-ph].
\bibitem{7} Y. Huang, J. He, H. F. Zhang and X. R. Chen, J. Phys. G {\bf 41}, no.11, 115004 (2014);\\                      arXiv:1305.4434 [nucl-th].
\bibitem{8} R. Aaij {\it et al.} (LHCb Collaboration), Phys. Rev. Lett. {\bf 115},
                                  072001 (2015);\\ arXiv:1507.03414 [hep-ex].
\bibitem{9} R. Aaij {\it et al.} (LHCb Collaboration), Phys. Rev. Lett. {\bf 122},
                                  222001 (2019);\\ arXiv:1904.03947 [hep-ex].
\bibitem{10} X.-Y. Wang, X.-R. Chen, and J. He, Phys. Rev. D {\bf 99}, 114007 (2019).
\bibitem{11} J. He, Eur. Phys. J. C {\bf 79}, 393 (2019);\\
                            arXiv:1903.11872 [hep-ph].
\bibitem{12} C.-J. Xiao {\it et al.}, Phys. Rev. D {\bf 100}, 014022 (2019).
\bibitem{13} A. Ali {\it et al.}, arXiv:1907.06507 [hep-ph].
\bibitem{14} H. X. Chen, W. Chen and S.-L. Zhu, Phys. Rev. D {\bf 100}, 051501 (2019);\\
                                 arXiv:1903.11001 [hep-ph].
\bibitem{15} R. Chen, Z. F. Sun, X. Liu and S.-L. Zhu, Phys. Rev. D {\bf 100}, 011502 (2019);\\
                                     arXiv:1903.11013 [hep-ph].
\bibitem{16} F. K. Guo, H. J. Jing, U. G. Meissner and S. Sakai, Phys. Rev. D {\bf 99}, 091501 (2019);\\
                                    arXiv:1903.11503 [hep-ph].
\bibitem{17} M. Z. Liu {\it et al.}, Phys. Rev. Lett. {\bf 122}, 242001 (2019);\\
                                 arXiv:1903.11560 [hep-ph].
\bibitem{18} J. R. Zhang, Eur. Phys. J. C {\bf 79}, 1001 (2019);\\
                           arXiv:1904.10711 [hep-ph].
\bibitem{19} H. Huang, J. He and J. Ping, arXiv:1904.00221 [hep-ph].
\bibitem{20} Y. Shimizu, Y. Yamaguchi and M. Harada, arXiv:1904.00587 [hep-ph].
\bibitem{21} C. W. Xiao, J. Nieves and E. Oset, Phys. Rev. D {\bf 100}, 014021 (2019);\\
                                arXiv:1904.01296 [hep-ph].
\bibitem{22} L. Meng, B. Wang, G. J. Wang and S.-L. Zhu, Phys. Rev. D {\bf 100}, 014031 (2019);\\
                                  arXiv:1905.04113 [hep-ph].
\bibitem{23} J. B. Cheng and Y. R. Liu, Phys. Rev. D {\bf 100}, 054002 (2019);\\
                                    arXiv:1905.08605 [hep-ph].
\bibitem{24} R. Aaij {\it et al.} (LHCb Collaboration), Sci. Bull. {\bf 66},1278 (2021);\\
                                  arXiv:2012.10380 [hep-ex].
\bibitem{25} M. Z. Liu, Y. W. Pan and L. S. Geng, Phys. Rev. D {\bf 103}, 034003 (2021);\\
                                    arXiv:2011.07935 [hep-ph].
\bibitem{26} Y. Irie, M. Oka and S. Yasui, Phys. Rev. D {\bf 97}, 034006 (2018);\\
                                    arXiv:1707.04544 [hep-ph].
\bibitem{27} V. V. Anisovich {\it et al.}, Int. J. Mod. Phys. A {\bf 30}, 1550190 (2015);\\
                                    arXiv:1509.04898 [hep-ph].
\bibitem{28} Z. G. Wang, Eur. Phys. J. C {\bf 76}, 142 (2016);\\
                                    arXiv:1509.06436 [hep-ph].
\bibitem{29} R. Chen, J. He and X. Liu, Chin. Phys. C {\bf 41}, 103105 (2017);\\
                                     arXiv:1609.03235 [hep-ph].
\bibitem{30} T. Gutsche and V. E. Lyubovitskij, Phys. Rev. D {\bf 100}, 094031 (2019);\\
                                     arXiv:1910.03984 [hep-ph].
\bibitem{31} H. X. Chen, L. S. Geng, W. H. Liang {\it et al.}, Phys. Rev. C {\bf 93}, 065203 (2016);\\
                                 arXiv:1510.01803 [hep-ph].
\bibitem{32} C. W. Shen, H. J. Jing, F. K. Guo and J. J. Wu, arXiv:2008.09082 [hep-ph].
\bibitem{33} A. Feijoo, V. K. Magas, A. Ramos and E. Oset, Eur. Phys. J. C {\bf 76}, no.8, 446 (2016);\\
                           arXiv:1512.08152 [hep-ph].
\bibitem{34} J. X. Lu, E. Wang, J. J. Xie, L. S. Geng and E. Oset, Phys. Rev. D {\bf 93}, 094009 (2016);\\
                           arXiv:1601.00075 [hep-ph].
\bibitem{35} W. Y. Liu {\it et al.}, Phys. Rev. D {\bf 103}, 034019 (2021);\\
                           arXiv:2012.01804 [hep-ph].
\bibitem{36} M. Ablikim {\it et al.} (BESIII Collaboration), Phys. Rev. Lett. {\bf 126}, 102001 (2021);\\
                           arXiv:2011.07855 [hep-ex].
\bibitem{37} L. Meng, B. Wang and S. L. Zhu, Phys. Rev. D {\bf 102}, 111502 (2020);\\
                                    arXiv:2011.08656 [hep-ph].
\bibitem{38} Z. Yang {\it et al.}, Phys. Rev. D {\bf 103}, 074029 (2021);\\
                           arXiv:2011.08725 [hep-ph].
\bibitem{39} F. L. Wang, R. Chen and X. Liu, Phys. Rev. D {\bf 103}, 034014 (2021);\\
\bibitem{40} F. L. Wang, X. D. Yang, R. Chen and X. Liu, Phys. Rev. D {\bf 103}, 054025 (2021);\\
                           arXiv:2101.11200 [hep-ph].
\bibitem{41} H. Ohnishi, F. Sakuma, T. Takahashi, Prog. Part. Nucl. Phys. {\bf 113}, 103773 (2020);\\
                                  arXiv:1912.02380 [nucl-ex].
\bibitem{42} K. Aoki {\it et al.}, arXiv:2110.04462 [nucl-ex].
\bibitem{43} S. Clymton, H.-J. Kim and H.-C. Kim, Phys. Rev. D {\bf 104}, 014023 (2021);\\
                                  arXiv:2102.08737 [hep-ph].
\bibitem{44}  E. Ya. Paryev and Yu.T. Kiselev, Nucl. Phys. A {\bf 978}, 201 (2018);\\
                                     arXiv:1810.01715 [nucl-th].
\bibitem{45}  A. Sibirtsev and W. Cassing, Nucl. Phys. A {\bf 641}, 476 (1998);\\
                                    arXiv:nucl-th/9805021.
\bibitem{46}  E. Ya. Paryev and Yu.T. Kiselev, Phys. At. Nucl. {\bf 80} (1), 67 (2017);\\
                                    arXiv:1510.00155 [nucl-th].
\bibitem{47}  E. Ya. Paryev, Yu. T. Kiselev and Yu. M. Zaitsev, Nucl. Phys. A {\bf 968}, 1 (2017).
\bibitem{48}  E. Ya. Paryev, Nucl. Phys. A {\bf 1007}, 122133 (2021);\\
                                    arXiv:2102.00789 [nucl-th].
\bibitem{49}  E. Ya. Paryev, Nucl. Phys. A {\bf 1013}, 122222 (2021);\\
                                    arXiv:2106.00353 [nucl-th].
\bibitem{50} S. V. Efremov and E. Ya. Paryev, Eur. Phys. J. A {\bf 1}, 99 (1998).
\bibitem{51} E. Ya. Paryev, Eur. Phys. J. A {\bf 7}, 127 (2000).
\bibitem{52} E. Ya. Paryev, Eur. Phys. J. A {\bf 9}, 521 (2000).
\bibitem{53} E. Ya. Paryev, Chinese Physics C, Vol. {\bf 42}, No. (8), 084101 (2018).
\bibitem{54} R. L. Anderson {\it et al.}, Phys. Rev. Lett. {\bf 38}, 263 (1977).
\bibitem{55} G. R. Farrar {\it et al.}, Phys. Rev. Lett. {\bf 64}, 2996 (1990).
\bibitem{56} V. Flaminio {\it et al.}, Compilation of Cross Sections.\\
             II: $K^+$ and $K^-$ Induced Reactions. CERN-HERA {\bf 83-02}, (1983).
\bibitem{57} C. Gobbi, C. B. Dover and A. Gal, Phys. Rev. C {\bf 50}, 1594 (1994).
\bibitem{58} G. W. London {\it et al.}, Phys. Rev. {\bf 143}, 1034 (1966).
\bibitem{59} A. Ali {\it et al.} (The GlueX Collaboration), Phys. Rev. Lett. {\bf 123}, 072001 (2019);\\
                                    arXiv:1905.10811 [nucl-ex].
\bibitem{60} E. Ya. Paryev, Chinese Physics C, Vol. {\bf 44}, No. (10), 104101 (2020);\\
                                    arXiv:2007.01172 [nucl-th].
\bibitem{61} J. Adamczewski-Musch {\it et al.} (HADES Collaboration with PANDA@HADES Collaboration),\\
                                      Eur. Phys. J. A {\bf 57}, 138 (2021); arXiv:2010.06961 [nucl-ex].
\bibitem{62} J. S. Lindsey and G. A. Smith, Phys. Rev. {\bf 147}, 913 (1966).
\bibitem{63} H. X. Chen, W. Chen, X. Liu, and X. H. Liu, Eur. Phys. J. C {\bf 81}, 409 (2021);\\
                                     arXiv:2011.01079 [hep-ph].
\bibitem{64} Z. G. Wang, Int. J. Mod. Phys. A {\bf 36}, 2150071 (2021);\\
                                     arXiv:2011.05102 [hep-ph].
\bibitem{65} U. Ozdem, Eur. Phys. J. C {\bf 81}, 277 (2021);\\
                                      arXiv:2102.01996 [hep-ph].
\bibitem{66} B. Wang, L. Meng, and S. L. Zhu, Phys. Rev. D {\bf 101}, 034018 (2020);\\
                                     arXiv:1912.12592 [hep-ph].
\bibitem{67} R. Chen, Phys. Rev. D {\bf 103}, 054007 (2021); [arXiv:2011.07214 [hep-ph]];\\
             R. Chen, Eur. Phys. J. C {\bf 81}, 122 (2021) [arXiv:2101.10614 [hep-ph]].
\bibitem{68} C. W. Xiao, J. Nieves, and E. Oset, Phys. Lett. B {\bf 799}, 135051 (2019);\\
                                     arXiv:1906.09010 [hep-ph].
\bibitem{69} C. W. Xiao, J. J. Wu, and B. S. Zou, Phys. Rev. D {\bf 103}, 054016 (2021);\\
                                     arXiv:2102.02607 [hep-ph].
\bibitem{70} Q. Wu, D.-Y. Chen, and R. Ji, Chin. Phys. Lett. {\bf 38}, 071301 (2021);\\
                                     arXiv:2103.05257 [hep-ph].
\bibitem{71} C. W. Shen, J. J. Wu, and B. S. Zou, Phys. Rev. D {\bf 100}, 056006 (2019);\\
                                     arXiv:1906.03896 [hep-ph].
\bibitem{72} E. Santopinto and A. Giachino, Phys. Rev. D {\bf 96}, 014014 (2017);\\
                                     arXiv:1604.03769 [hep-ph].
\bibitem{73} K. Azizi, Y. Sarac, and H. Sundu, Phys. Rev. D {\bf 103}, 094033 (2021);\\
                                     arXiv:2101.07850 [hep-ph].
\bibitem{74} F. Z. Peng, M. J. Yan, M. Sanchez Sanchez, and M. P. Valderrama, Eur. Phys. J. C {\bf 81}, 666 (2021);\\
                                      arXiv:2011.01915 [hep-ph].
\bibitem{75} V. Kubarovsky and M. B. Voloshin, Phys. Rev. D {\bf 92}, 031502 (2015);\\
                                     arXiv:1508.00888 [hep-ph].
\bibitem{76} V. Kubarovsky and M. B. Voloshin, arXiv:1609.00050 [hep-ph].
\bibitem{77} M. Karliner and J. L. Rosner, Phys. Lett. B {\bf 752}, 329 (2016);\\
                                    arXiv:1508.01496 [hep-ph].
\bibitem{78} X.-Y. Wang, J. He, X.-R. Chen, Q. Wang, and X. Zhu, Phys. Lett. B {\bf 797}, 134862 (2019);\\
                                     arXiv:1906.04044 [hep-ph].
\bibitem{79} X.-Y. Wang, J. He, and X. Chen, Phys. Rev. D {\bf 101}, 034032 (2020);\\
                                     arXiv:1912.07156 [hep-ph].
\bibitem{80} Ye. Golubeva, W. Cassing, and L. A. Kondratyuk, Eur. Phys. J. A {\bf 14}, 255 (2002);\\
                                      arXiv:nucl-th/0202084.
\bibitem{81} S. Acharya {\it et al.} (ALICE Collaboration), arXiv:2201.05352 [nucl-ex].
\bibitem{82} S. Aoki, Nucl. Phys. A {\bf 644}, 365 (1998).
\bibitem{83} F. Ferro {\it et al.}, Nucl. Phys. A {\bf 789}, 209 (2007).
\bibitem{84} E. Ya. Paryev, Nucl. Phys. A {\bf 1017}, 122352 (2022);\\
                                    arXiv:2111.14101 [nucl-th].
\bibitem{85} Ye. S. Golubeva {\it et al.}, Eur. Phys. J. A {\bf 17}, 275 (2003);\\
                                      arXiv:nucl-th/0212074.
\bibitem{86} J. M. Hauptman, J. A. Kadyk, and G. H. Trilling, Nucl. Phys. B {\bf 125}, 29 (1977).
\bibitem{87} Y. Nara {\it et al.}, Nucl. Phys. A {\bf 614}, 433 (1997).
\bibitem{88} B. G. Yu, T. K. Choi, and C. R. Ji, J. Phys. G {\bf 32}, 387 (2006);\\
                                     arXiv:nucl-th/0408006.
\bibitem{89} A. Faessler {\it et al.}, Phys. Rev. C {\bf 70}, 035211 (2004);\\
                                     arXiv:nucl-th/0407075.
\bibitem{90} A. N. Hiller Blin {\it et al.}, Phys. Rev. D {\bf 94}, 034002 (2016);\\
                                     arXiv:1606.08912 [hep-ph].
\bibitem{91} D. Winney {\it et al.}, Phys. Rev. D {\bf 100}, 034019 (2019);\\
                                     arXiv:1907.09393 [hep-ph].
\bibitem{92} E. Ya. Paryev, Nucl. Phys. A {\bf 996}, 121711 (2020);\\
                                     arXiv:2003.00788 [nucl-th].
\bibitem{93} C. Amsler {\it et al.}, Phys. Lett. B {\bf 667}, 1 (2008).

\end{thebibliography}
\end{document}